\documentclass{emulateapj}
\usepackage{epsfig}
\usepackage{lscape}

\newcommand{\etal}{\mbox{et al.}}
\newcommand{\ergcms}{\mbox{erg cm$^{-2}$ s$^{-1}$}}

\newcommand{\ergsec}{\mbox{erg s$^{-1}$}}

\newcommand{\phcms}{\mbox{photons cm$^{-2}$ s$^{-1}$}}
\newcommand{\degree}{$^\circ$}

\newcommand{\chandra}{{\it Chandra}}

\newcommand{\xmm}{{\it XMM-Newton}}

\newcommand{\sgrastar}{\mbox{Sgr A$^*$}}
\newcommand{\program}[1]{{\tt {#1}}}

\shortauthors{Muno \etal}
\shorttitle{Galactic Center X-ray Sources}

\begin{document}
\title{A Catalog of X-ray Point Sources from Two Megaseconds of Chandra Observations of the Galactic Center}
\author{
M. P. Muno,\altaffilmark{1} 
F. E. Bauer,\altaffilmark{2}
%R. G. Arendt,\altaffilmark{6}
F. K. Baganoff,\altaffilmark{3} 
R. M. Bandyopadhyay,\altaffilmark{4}
G. C. Bower,\altaffilmark{5}
W. N. Brandt,\altaffilmark{6} 
P. S. Broos, \altaffilmark{6}
A. Cotera,\altaffilmark{7}
S. S. Eikenberry,\altaffilmark{4}
G. P. Garmire,\altaffilmark{6}
S. D. Hyman,\altaffilmark{8}
N. E. Kassim,\altaffilmark{9}
C. C. Lang,\altaffilmark{10}
%T. N. LaRosa,\altaffilmark{13}
T. J. W. Lazio,\altaffilmark{9}
C. Law,\altaffilmark{11}
J. C. Mauerhan,\altaffilmark{12}
M. R. Morris,\altaffilmark{12}
T. Nagata,\altaffilmark{13}
S. Nishiyama,\altaffilmark{14} 
S. Park,\altaffilmark{6}
%E. Pfahl,\altaffilmark{17}
S. V. Ram\`{i}rez,\altaffilmark{15}
S. R. Stolovy,\altaffilmark{15}
R. Wijnands,\altaffilmark{11}
Q. D. Wang,\altaffilmark{16}
Z. Wang,\altaffilmark{17}
and F. Yusef-Zadeh\altaffilmark{18}
}

\altaffiltext{1}{Space Radiation Laboratory, California Institute of 
Technology, Pasadena, CA 91104; mmuno@srl.caltech.edu}
\altaffiltext{2}{Columbia Astrophysical Laboratory, New York, NY, 10027, feb@astro.columbia.edu}
%\altaffiltext{6}{Observational Cosmology Laboratory, Code 665, Goddard Space Flight Center, Greenbelt, MD 20771}
\altaffiltext{3}{Kavli Institute for Astrophysics and Space Research,
Massachusetts Institute of Technology, Cambridge, MA 02139}
\altaffiltext{4}{Department of Astronomy, University of Florida, Gainesville, FL 32611} 
\altaffiltext{5}{Department of Astronomy and Radio Astronomy Laboratory, University of California, Berkeley, CA 94720-3411}
\altaffiltext{6}{Department of Astronomy and Astrophysics, 
The Pennsylvania State University, University Park, PA 16802}
\altaffiltext{7}{SETI Institute, 515 N. Whisman Rd., Mountain View, CA}
\altaffiltext{8}{Department of Physics and Engineering, Sweet Briar College, Sweet Briar, VA 24595}
\altaffiltext{9}{Remote Sensing Division, Naval Research Laboratory, Washington, DC 20375-5351}
\altaffiltext{10}{Department of Physics and Astronomy, University of Iowa, Van Allen Hall, Iowa City, IA 52242}
\altaffiltext{11}{Astronomical Institute ``Anton Pannekoek'', University of Amsterdam, Kruislaan 403, 1098 SJ, The Netherlands}
\altaffiltext{12}{Department of Physics and Astronomy, University of California, Los Angeles, CA 90095}
%\altaffiltext{12}{Department of Biological \& Physical Sciences, Kennesaw State University, 1000 Chastain Road, Kennesaw, GA 30144}
\altaffiltext{13}{Department of Astronomy, Kyoto University, Kyoto 606-8502, Japan}
\altaffiltext{14}{National Astronomical Observatory of Japan (NAOJ), National Institutes of Natural Sciences (NINS), Tokyo 181-8588, Japan}
%\altaffiltext{15}{Kavli Institute for Theoretical Physics, University of California, Santa Barbara, CA 93106}
\altaffiltext{15}{Spitzer Science Center, California Institute of Technology, Pasadena, CA 91125}
\altaffiltext{16}{Astronomy Department, University of Massachusetts, Amherst, MA 01003, USA}
\altaffiltext{17}{Department of Physics, McGill University, Montreal, QC H3A 2T8, Canada}
\altaffiltext{18}{Department of Physics and Astronomy, Northwestern University, Evanston, IL 60208}

\begin{abstract}
We present a catalog of 9017 X-ray sources identified in \chandra\
observations of a 2$\times$0.8\degree\ field around the Galactic center.
This enlarges the number of known X-ray sources in the region by a factor
of 2.5. The catalog incorporates all of the ACIS-I observations as of 
2007 August, which total 2.25 Msec of exposure.
At the distance to the Galactic center (8~kpc), we are sensitive to sources
with luminosities of $4\times10^{32}$~\ergsec\ (0.5--8.0 keV; 90\% confidence) 
over an area of one square degree, and up to an order of magnitude more 
sensitive in the deepest exposure (1.0~Msec) around \sgrastar. 
The positions of 60\% of our sources are accurate to $<$1\arcsec 
(95\% confidence), and 20\% have positions accurate to $<$0\farcs5.
We search for variable sources, and find that 3\% exhibit flux variations 
within an observation, 10\% exhibit variations from observation-to-observation.
We also find one source, CXOUGC J174622.7--285218, with a periodic 
1745~s signal (1.4\% chance probability), which is probably a 
magnetically-accreting cataclysmic variable.
We compare the spatial distribution of X-ray sources to a model for the
stellar distribution, and find 2.8$\sigma$ evidence for 
excesses in the numbers of X-ray sources in the region of recent star 
formation encompassed by the Arches, Quintuplet, and Galactic center star
clusters. These excess sources are also seen in the luminosity distribution 
of the X-ray sources, which is flatter near the Arches and Quintuplet than 
elsewhere in the field. These excess point sources, along with a similar 
longitudinal asymmetry in the distribution of diffuse iron emission that 
has been reported by other authors, probably have their origin in the young 
stars that are prominent at $l$$\approx$0.1\degree.
\end{abstract}


\section{Introduction}

Stars are detectable as X-ray sources at several important stages of their
lives. Pre-main sequence stars are X-ray sources because of
their enhanced magnetic activity \citep{pf05}. Massive OB and Wolf-Rayet stars
produce X-rays through shocks in their stellar winds \citep{ber97,gagne05},
and possibly from magnetically-confined plasma close to their stellar surfaces
\citep{wc07}. Neutron
stars are bright X-ray sources if
they are young and still have latent heat from what was once the
stellar core \citep{wwn96}, 
if they accelerate particles in rotating,
moderate-strength ($B$$\sim$$10^{12}$ G) fields \citep{gs06}, or if they have
extremely strong fields ($B$$\sim$$10^{14}$ G) that decay and
accelerate particles \citep{wt06}.  
White dwarfs, neutron stars, and black holes are bright 
X-ray sources if they are accreting matter from a binary
companion \citep{war95,psa06}, or in principle from the interstellar medium 
\citep[see, e.g.,][]{per03}. 
Therefore, X-ray surveys can be used to study the life cycles of stars,  
particularly their start and end points.

Here we present a catalog of X-ray sources detected in \chandra\ observations
toward the inner 2\degree\ by 0.8\degree\ of the Galaxy. The region
encompasses about 1\% of the Galactic stellar mass \citep{lzm02}, 
and possibly up to 10\% of the Galactic population of young, massive stars 
\citep{mp79,fig04}. Therefore,
these data provide a statistically meaningful sample of the 
Galactic stellar population.

Previous catalogs based on \chandra\ data on the Galactic center
have been published by \citet{m-cat} using 630 ks of
data taken through 2002 June on the central
17\arcmin$\times$17\arcmin\ around \sgrastar, and by \citet{m-wide} using
observations taken through 2005 June on the inner 
2\degree$\times$0.8\degree\ of the Galaxy. However, since the publication
of these catalogs, a large amount of new
data have been obtained. These data increase the
number of point sources identified by a factor of 2.5. They also provide much 
better astrometry for individual X-ray sources. 
The improvement in astrometry enables the identification of rare objects such 
as Wolf-Rayet stars, X-ray
binaries, and rotation-powered pulsars, through comparisons of our 
X-ray catalog with radio and infrared data sets (e.g., Mauerhan, J. 
\etal, in prep). Therefore, we provide 
here an updated catalog of point sources, which incorporates and 
supercedes the previous catalogs. We also describe the spatial and luminosity
distributions of the X-ray sources. 

Throughout this paper, we adopt 
a distance to the Galactic center of $D$=8~kpc \citep{reid93,mcn00}, 
and an average absorption
column of $N_{\rm H}$=$6\times10^{22}$~cm$^{-2}$ \citep{bag03}.

\section{Observations\label{sec:obs}}

As of 2007 August, the central 2\degree$\times$0.8\degree\ of the
Milky Way has been observed with the imaging array of the \chandra\
Advanced CCD Imaging Spectrometer \citep[ACIS-I;][]{wei02}\footnote{See
also http://cxc.harvard.edu/proposer/POG/html/ACIS.html} on numerous
occasions. The majority of the new sources in this catalog come from 
600 ks of exposure that we obtained in fifteen 40 ks pointings covering
$\approx$1\degree\ of the Galactic center. The new data since 2005
also includes 370 ks on the central 20 pc around \sgrastar, 100~ks on
the Arches cluster \citep{wdl06}, and 100~ks on Sgr C. We also include
sources previously identified in 630 ks of data on the inner 20 pc
around \sgrastar\ \citep{m-cat, m-ps}; thirty 12 ks exposures of the
2\degree$\times$0.8\degree\ survey obtained by 
\citet[][see also Muno \etal\ 2006a]{wgl02}\nocite{m-wide}; and
deep pointings toward the Radio Arches \citep[50 ks;][]{lyz04} and Sgr
B2 \citep[100 ks;][]{tak02}.  The dates, observation identifiers
(ObsIds), durations, locations, roll angles, and some values relevant for the
astrometry (\S2.1) for each exposure are
listed in Table~\ref{tab:obs}. The observations in the table are 
sorted by right ascension and declination, so that observations near
the same point are grouped.

\begin{figure*}[htp]
\centerline{\epsfig{file=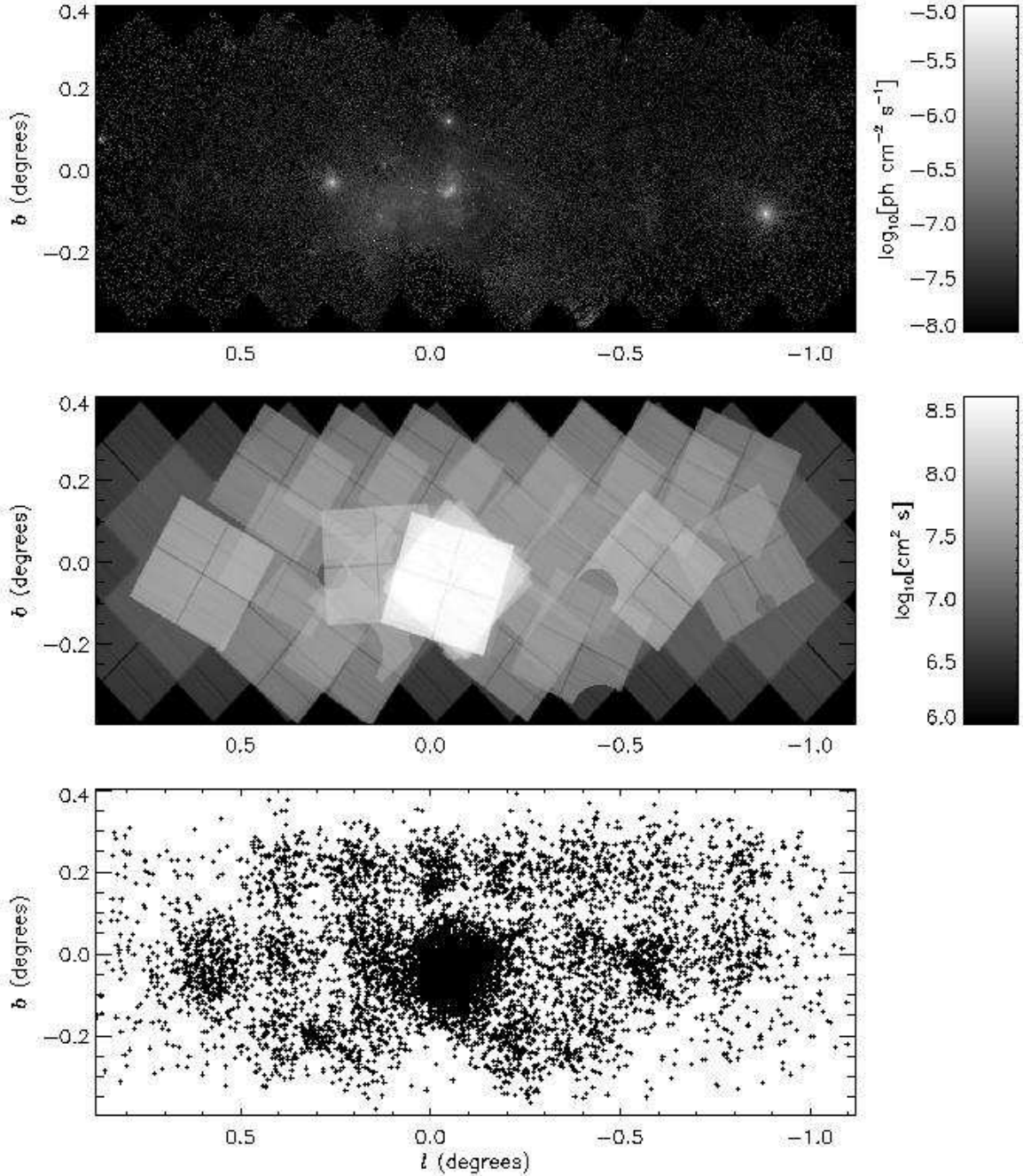,width=0.9\linewidth}}
\caption{
Basic results from the survey. The {\it top panel} contains a
composite image of the field, in which the counts have been divided by
an exposure map to provide an estimate of the 0.5-8.0 keV flux.  The
{\it middle panel} contains the exposure map for an energy of 4~keV,
in units of the product of the effective area times the exposure
time. Some holes are visible where we have excluded regions where
bright transients and their associated dust-scattering halos were
present in some individual observations, because these degraded the
sensitivity.  The {\it bottom panel} illustrates the locations of
point sources in our sample. The regions with the largest exposure
have the greatest concentrations of point sources.  }
\label{fig:obs}
\end{figure*}

\begin{figure*}[htp]
\centerline{\epsfig{file=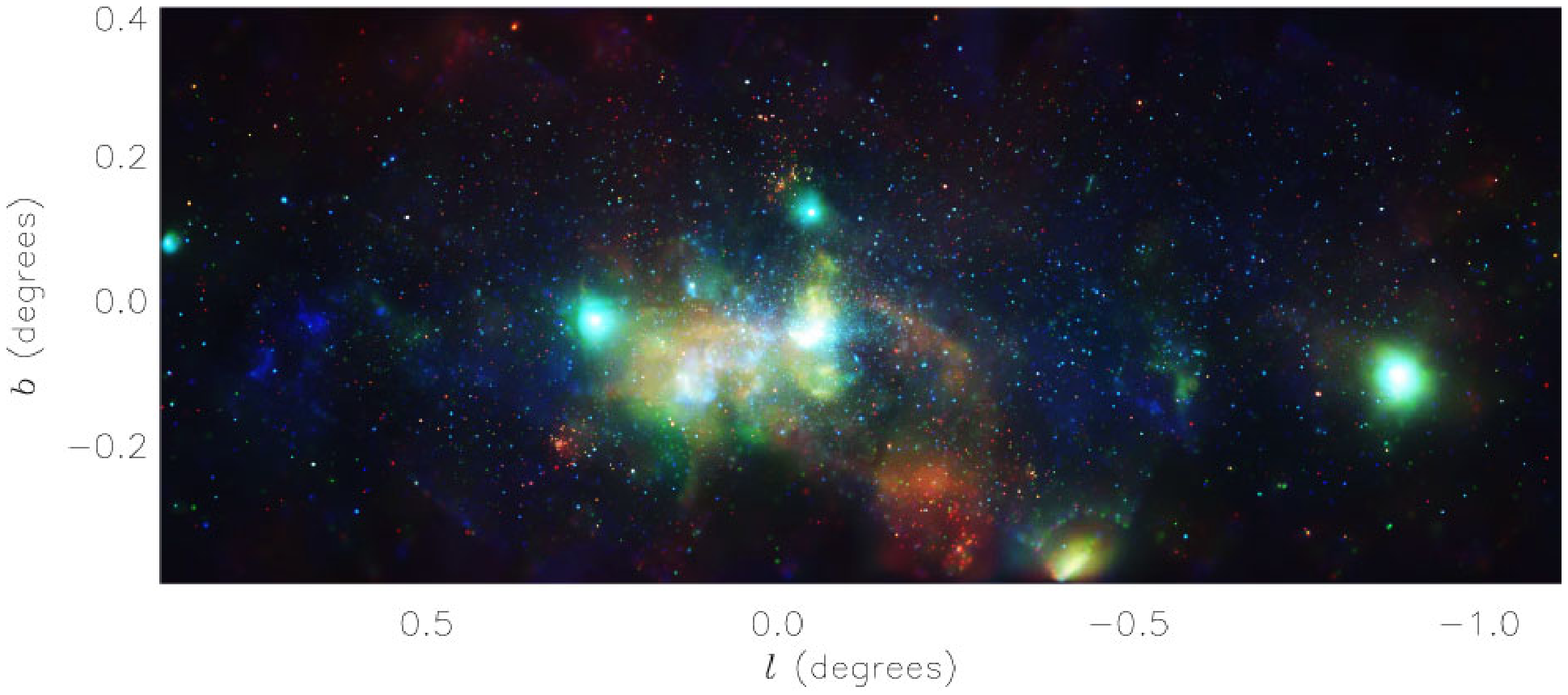,width=0.95\linewidth}}
\caption{
Three-color image of the survey area. Red is 1--3~keV, 
green is 3--5~keV, and blue is 5--8~keV. Each band was 
adaptively smoothed using the CIAO tool {\tt csmooth}, 
and then normalized using an exposure map. Some artifacts
can be seen at the boundaries of chip edges, particularly near 
where bright, transient X-ray sources appeared.
}
\label{fig:rgb}
\end{figure*}

The ACIS-I is a set of four, 1024-by-1024 pixel CCDs, covering a field
of view of 17\arcmin\ by 17\arcmin. When placed on-axis at the focal
plane of the grazing-incidence X-ray mirrors, the imaging resolution
is determined primarily by the pixel size of the CCDs, 0\farcs492.
The CCD frames are read out every 3.2~s, which provides the nominal
time resolution of the data.  The CCDs also measure the energies of
incident photons within a calibrated energy band of 0.5--8~keV, with a
resolution of 50--300 eV (depending on photon energy and distance from
the read-out node). However in some of the earlier, shallow exposures
(ObsIDs 2267 through 2296), an event filter was employed on the
satellite that removed X-rays with energies below 1 keV before the
data were sent to the ground. The lack of 0.5--1.0 keV photons had
a minor impact on our results, because there were only 76 sources
that were detected below 
2~keV for which the photometry was derived entirely from the ObsIDs
2267 through 2296.\footnote{For these sources, we 
under-estimate the flux by $\approx$25\%. The soft color is also 
systematically high. For instance, sources
with $N_{\rm H} \approx 10^{21}$ cm$^{-2}$ will have $HR0$$\approx$$-0.5$ 
using a 0.5--2.0 keV soft band, and $HR0$$\approx$$-0.3$ using a 1.0--2.0 keV
soft band.} We omitted ObsID 242 from 
our analysis, because it was taken with the detector at a cooler 
temperature (110 K, versus 120 K). A flux image and composite exposure
map is displayed in Figure~\ref{fig:obs}, and an adaptively-smoothed
three-color image is displayed in Figure~\ref{fig:rgb}.

The data were processed using the 
\chandra\ Interactive Analysis of Observations 
(CIAO)\footnote{http://cxc.harvard.edu/ciao/} package. 
The data were processed as they arrived, so we used CIAO versions 
3.3 and 3.4. Information on the detectors was taken from the Calibration
Database (CALDB)\footnote{http://cxc.harvard.edu/caldb/}
versions 3.2.1 and 3.3.0. The differences between the
two versions of the software were too minor to justify re-processing
the older portions of the dataset. We only used data from ACIS-I; data from
the S array was omitted because was offset far from the aim point, and 
the large point spread function on the detector resulted in bad stellar
confusion.

We started with the level 1 event files provided by the \chandra\
X-ray Center (CXC), and reprocessed the data using the tool {\tt
acis\_process\_events} in order to remove pixel randomization and
apply more recent energy calibration.  We then removed events
associated with bad pixels, and applied the standard grade filters to
the events and good-time filters supplied by the CXC.  We applied {\tt
acis\_run\_hotpix} to flag events associated with cosmic-rays, and
removed them from the event list (We did not run {\tt
acis\_detect\_afterglow}, because it sometimes removes genuine X-ray
events. We did, however, later remove sources that were cosmic ray 
afterglows; see \S2.2).  We then searched each observation for time 
intervals when
particles encountering the detector caused the background event rate
to flare to $\ge$3$\sigma$ above the mean level, and removed
them. These background flares were found in 12 observations, and
lasted $<$5\% of the duration of each observation.  Next, we applied
the sub-pixel event repositioning algorithm of \citet{li04}.  Finally,
if an astrometric correction was available from the CXC for any
observation, we applied it at this point, by modifying the header
keywords for the event file, and by correcting the columns for the
right ascension and declination in the aspect solution provided with
the observation.

Before proceeding to explain our algorithms for source detection, we
would like to explain some minor weaknesses of our approach.
Unfortuantely, because the data were searched for sources as they
arrived, and because the exposures were highly non-uniform across the
field, some parameters of our detection algorithm, particularly the
detection threshholds, were not kept consistent. To compensate for
this, we did two things. First, we verified the reality of each source
as part of our photometric algorithm (\S2.2). This should eliminate
most spurious sources on the faint end, in a uniform manner. Second,
we determined the completeness limits of our survey using Monte Carlo
simulations that mimicked our source detection algorithms
(\S2.3). This is the best way to establish what portion of our sample
is complete.  With the experience we have gained, in principle we
could develop a more streamlined and straightforward approach to
building the inital catalog. However, it would take several months of
computer time to reprocess the data, or a similar amount of time
re-writing our software to be more efficient.  The improvements in the
final catalog would be slight, so we decided not to delay releasing
our catalog any further.

We are making the data products from the following sections available
in FITS format from a web 
site.\footnote{http://www.srl.caltech.edu/gc\_project/xray.html}
The catalog itself will also be available with the electronic version
of the paper.

\subsection{Source Detection and Initial Localization}

Source detection and localization was approached iteratively.
We first searched for point sources in each observation separately. The 
locations of the point sources found in the first stage were used 
to refine the astrometry. Second, the astrometrically-corrected images 
were combined to search for fainter point sources. Finally, the source
lists from the individual observations were merged with those from the 
combined images.

We searched each observation individually for point sources
using the wavelet decomposition algorithm {\tt wavdetect}
\citep{free02}.  We
employed the default ``Mexican Hat'' wavelet, and used a sensitivity
threshold of $10^{-7}$. This threshold 
roughly corresponds to the chance of detecting a
spurious source in an area corresponding to the point spread function (PSF), 
if the local background is spatially
uniform. For the earlier data, taken before 2006, we used images 
at three different
resolutions: one at 0\farcs5 resolution covering the inner
1024$\times$1024 pixels, one at 1\arcsec\ resolution covering the
inner 2048$\times$2048 pixels, and one at 2\arcsec\ resolution
covering the entire field. For later observations, we simplified the
process and used only two resolutions: 0\farcs5 covering the inner
2048$\times$2048 pixels, and 2\arcsec\ covering the entire
field. Using a test field, we confirmed that there was no difference
in the number of sources detected using the two techniques; the only
difference is that the technique that used three, smaller
images was computationally faster. We used wavelet scales that
increased by a factor of $\sqrt{2}$, over the range of 1--4 pixels for 
the 0\farcs5 image, 1--8 pixels
for the 1\arcsec\ image, and 1--16 pixels for the 2\arcsec\ image.  For each
resolution, we made images in three energy bands: 0.5--8.0~keV to
cover the full bandpass, 0.5--2.0~keV to provide sensitivity to
foreground sources, and 4--8 keV to provide sensitivity to
highly-absorbed sources. For each image, a matching exposure map was 
generated for photons with an energy of 4~keV, so that the 
wavelet algorithm could keep track of regions with rapidly-varying
exposure, such as bad columns and the edges of the CCDs.

The lists derived from each image resolution were combined to form
master source lists for each energy band. We found that the
positions would be most accurate from the sources identified in the
image with the finest resolution. Therefore, we discarded sources from
the lower-resolution images if their separations from sources
identified at high resolution were smaller than the radii of the 90\%
contour of the PSF. In this way, we produced three
lists for each observation, one for each energy band. 

Next, we used the point sources 
detected so far to register the absolute
astrometry to the Two Micron All-Sky Survey \citep[2MASS;][]{skr06}. 
The 2MASS frame is consistent with the International Celestrial Reference 
System to within 15 mas. We compared the positions of 2MASS sources to 
those of X-ray sources detected in the 0.5--2.0 keV band that were 
identified by {\tt wavdetect} as having $>$3$\sigma$ significance, and 
identified matches as those with offsets $<$1\arcsec. The offsets between
the 2MASS and \chandra\ frames were computed using a least-chi-squared
algorithm. The X-ray sources that we used for astrometry are flagged in 
Table~\ref{tab:positions} (see \S\ref{sec:tdesc}).

For observations longer than 20
ks, we found between 3 and 36 X-ray sources in the soft band
that could be associated unambiguously with stars in the 2MASS catalog, 
and so we used the average offsets between the X-ray and infrared sources to
correct the astrometry of the X-ray observations. We evaluated the accuracy 
of the registration based on the standard deviation in the mean
of the offsets of the individual stars. The registraton was accurate
to 0\farcs06 for the deepest exposures, and to 0\farcs2
for the shallower ones (1$\sigma$). Unfortunately, for exposures shorter than
20~ks, too few X-ray sources were found with 2MASS counterparts to
correct the astrometry to better than the default value, 0\farcs5
(1$\sigma$).

Once the deepest observation at each point was registered to the 2MASS
frame, the shallower observations were registered using the offsets of
X-ray sources detected in the 0.5--8.0 keV band in pairs of
observations. Between 2 and 759 X-ray sources matched between the deepest
and shallower observations, depending upon the exposure time of the
shallower observation.  The uncertainty in the
astrometry of each observation is listed in the last column of
Table~\ref{tab:obs}. The composite image and exposure map for our 
survey are displayed in Figure~\ref{fig:obs}.

Having corrected the astrometry for fields that included deep 
observations, we then combined subsets of the images in order 
to perform a deeper search for point sources. Two wavelet algorithms
were used, on the series of images listed below.
First, the tool {\tt wavdetect} \citep{free02} was used to identify point 
sources in:
\begin{itemize}
\item Composite images made from all of the observations of \sgrastar. 
Twelve, 1024$\times$1024 images were made, in four resolutions
(0\farcs25, 0\farcs5, 1\arcsec, and 2\arcsec) and using three energy 
bands for each resolution (0.5--8.0~keV, 0.5--2.0~keV, and 4--8~keV).
\item Three sets of composite images made from observations of \sgrastar\
in 2002, 2004, and 2005. These images were designed to be sensitive to 
faint, variable sources. The same image resolutions and energy bands 
were used as for the composite image of all of the \sgrastar\ data.
\item Composite images made from three pointings that were taken with 
same same roll angle, because the original 40~ks exposure had to be split
up to accomodate scheduling constraints (ObsIDs 7038, 7041, and 7042). 
Three images were made for 
each aimpoint, one for each of the 0.5--8.0~keV, 0.5--2.0~keV, and 4--8~keV
energy bands. Each image was made at 0\farcs5 resolution, and had
2048$\times$2048 pixels. 
\end{itemize}
The parameters used with {\tt wavdetect} were the same as for the 
individual observations. 

Second, we used the tools {\tt wvdecomp} and {\tt findpeak} in
the {\it zhtools} package written by A.\ Vikhlinin\footnote{{\tt
http://hea-www.harvard.edu/RD/zhtools}} to search for faint sources 
that fell below the {\tt wavdetect} threshold. 
We searched on wavelet 
scales of 1--3 pixels, and required that a candidate source be identified 
with a minimum signal-to-noise of 4.5, corresponding to 16 spurious sources
per 2048 by 2048 pixel image.
Five iterations of the search procedure were performed.
The tool {\tt wvdecomp} 
iteratively cleans the image of point sources identified in previous 
passes through the data, so it is more efficient at separating close
pairs of sources. Moreover, unlike {\tt wavdetect}, {\tt wvdecomp} 
does not use any information about the 
shape of the point spread function in searching for sources, so it is 
better at identifying point sources when observations 
with very different aimpoints have been combined. Therefore, we used
{\tt wvdecomp} on composite images generated from all data covering
the positions at which deep observations were obtained
(i.e., ObsIDs 3392, 4500, 5892, 7034--7048, and 944). Each image
was produced with 2048$\times$2048 pixels at 0\farcs5 resolution 
for the 0.5--8.0 keV band. 

We then generated a list containing the unique sources, by merging the 
lists generated by {\tt wavdetect} from individual observations and
from combined images, and from the lists generated by {\tt wvdecomp}. 
We found that almost all the duplicates could be removed by identifying
sources with separations smaller than the prescription for positional 
uncertainties in \citet{brandt01}: for sources offset from the center 
of each image by $\theta$$<$5\arcmin, the separation was cut at 0\farcs6, 
whereas for larger offsets it was cut at 
$0\farcs6 + (\theta - 5.0)/8.75$ ($\theta$ is in 
arcmin).\footnote{We use a different prescription 
in \S2.4 for the uncertainties on the positions in the catalog.}
For each image, we gave preference to positions from the
full-band sources, then from the soft sources, and finally from the hard
sources. Across observations, the priority was given to the deeper 
observations, and to sources detected with {\tt wavdetect} over those
detected with {\tt wvdecomp}. We examined the final list visually by
comparing it to images of the survey fields, and we removed several
hundred sources that were portions of extended, diffuse features
\citep[from][]{m-pwn}, and a 
couple dozen duplicates that were not identified automatically.  Finally, two
sources were not picked up by the detection algorithms because they 
were blended with nearby, brighter sources. We added these 
to our catalog by hand (CXOUGC J174502.8--282505 and J174617.4--281246).

At this stage, we considered our source lists to be provisional, both because
the search algorithms used non-uniform parameters, and because 
the large, spatially-variable background was likely to cause 
our wavelet algorithms to generate a significant number of spurious sources. 
In order to confirm their validity, we next computed photometry
for each provisional source.

\begin{figure}
\centerline{\epsfig{file=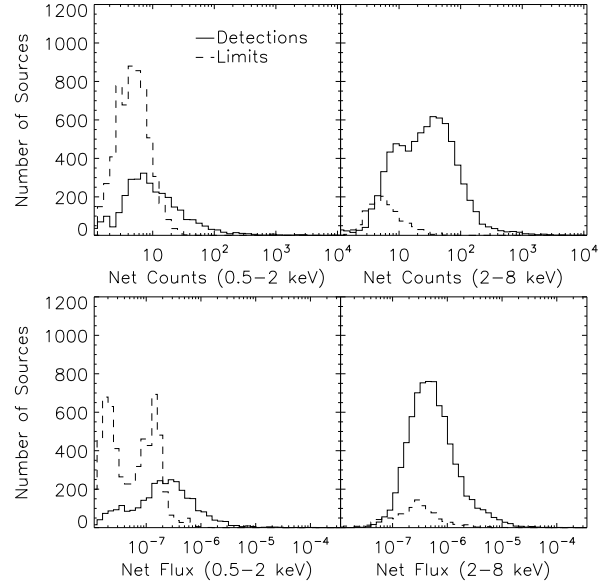,width=0.95\linewidth}}
\caption{{\it Top panels:} The distributions in net counts from individual
sources. No corrections were applied to account for the exposure across the 
survey, which varies by a factor of 10. Values for the 0.5--2.0 keV band
are plotted on the left, and for the 2--8 keV band on the right.  
{\it Bottom panels:} The distribution of fluxes (\phcms) from individual 
sources. The 0.5--2.0 keV~fluxes were derived by dividing the net count 
rates by the effective area and exposure the 0.5--2.0 keV band,
whereas the 2--8~keV fluxes were computed by dividing the counts into
into three energy bands (2.0--3.3~keV; 3.3--4.7~keV; and 4.7--8.0~keV), 
dividing by the respective effective areas and exposures, and 
summing the result. There are two peaks in each histogram, because 
the deeper observations were more sensitive to faint sources. In all 
panels, the solid lines are used for detections, and the dashed lines
are 90\% upper limits derived when a source was detected in one band,
but not the other.
}
\label{fig:lumdist}
\end{figure}

\subsection{Photometry}

We computed aperture photometry for each source using the 
\anchor{http://www.astro.psu.edu/xray/docs/TARA/ae\_users\_guide.html}{ACIS Extract} package, versions 3.96, 3.101, and 3.128 
\citep{broos02, tow03, get05}, along with some custom code.
The algorithm proceeded in several steps.

First, for each source and each observation, we obtained a model PSF 
with the CIAO tool {\tt mkpsf}. For most sources, we used a PSF for a
fiducial energy of 4.5 keV. However, if a source was only detected with 
{\tt wavdetect} in
the soft band, we used a PSF for an energy of 1.5 keV.  To determine a
region from which we would extract source counts, we then constructed
a polygon enclosing 90\% of the PSF. If the polygons for two sources 
overlapped in the observations in which the sources were closest to
the aim-point, we generated a smaller polygon. The final
extraction regions enclosed between 70\% and 90\% of the PSF. Sources
for which the PSF fraction was $<$90\% were considered to be
confused. Moreover, because the PSF grows rapidly beyond 7\arcmin\ from the 
aim-point, we also considered sources to be confused if they were located
beyond 7\arcmin\ from the aim-point and their PSFs overlapped. 
Photometry was not computed for observations in which confused sources
fell $>$7\arcmin off-axis. Fortunately, these second type of confused
sources were always located on-axis in another observation, or else
they would not have been identified. Finally, for similar reasons,
we only computed photometry for sources that lay within 7\arcmin\ of \sgrastar\
if the relevant observations had \sgrastar\ as the aim point.

Second, we extracted source event lists, source spectra, effective
area functions, and response matrices for each source in each
observation.  The detector responses and effective areas were obtained
using the CIAO tools {\tt mkacisrmf} and {\tt mkarf},
respectively. For each source, the spectra from all of the relevant
observations were summed. The responses and effective areas were
averaged, weighted by the exposures in each observation.

Third, we extracted background events from circular regions
surrounding each point source in each observation, omitting events
that fell within circles that circumscribed $\approx$90\% of the PSFs
around any point sources.  The background regions were chosen to
contain $\approx$100 total counts for the wide survey, and $\approx$1000
total counts for the deeper \sgrastar\ field. Fewer than 1\% of the
counts in the background region originate from known point
sources. For each source, the background spectra from all of the
relevant observations were scaled by the integrals of the exposure 
maps (in units of cm$^2$ s) over the source
and background regions, and then summed to create composite
background spectra.  

Fourth, we eliminated spurious sources. We compared the number of source and 
background counts 
to estimate the probability that no source was present, based on 
Poisson statistics \citep{wei07}. If a source had a $>$10\% chance of 
being spurious, we eliminated it from our catalog. 
We eliminated 1962 sources in this way. We also eliminated sources
in which the majority of events were cosmic ray afterglows. Specifically, 
we removed 46 sources because the events associated with the candidate source 
fell in a single pixel during 5--10 consecutive frames.  
 Our final catalog contains 9017 X-ray sources, 
and is listed in Table~\ref{tab:positions}. 
The majority of sources, 7152, were found with 
{\tt wavdetect}. Of the sources detected with {\tt wavdetect}, 
4823 were detected in the full band, 948 in the soft band, 
and 1381 in the hard band. Another 1865 sources were only detected with 
{\tt wvdecomp}. In the \sgrastar\ field alone, we found 3441 sources with 
{\tt wavdetect}, of which 
2715 were detected in the full composite image, 275 in 2002, 48 in 2004,
90 in 2005, and 313 in individual observations. An additional 
364 were found in the \sgrastar\ field with {\tt wvdecomp}.

Fifth, we compared the source and background spectra
using the Kolmogorov-Smirnov (KS) statistic, in order to flag
potentially-spurious objects that could be variations in the
background. Caution should be used when studying sources
that resemble the background. For instance, in the central parsec
around \sgrastar, there is an over-abundance of faint ($\la$$10^{-6}$
\phcms), soft point sources that have spectra consistent with that of the
background warm plasma (note that almost all of the excess bright
sources that we discuss in \S3.1 do have spectra that are distinct 
from the background). Therefore, we suspect that most of these are
$\sim$0.1~pc scale variations in the density of that
plasma. Unfortunately, we cannot be certain. Indeed, the spectrum of
the bright X-ray source associated with IRS~13
(CXOUGC~J174539.7--290029) resembles the background according to the KS
test. If the diffuse background is merely unresolved point sources
\citep{wgl02,rev06,rs07}, then most faint point sources should have
spectra that resemble the background. 

Sixth, we computed the net counts in the 
0.5--2.0~keV, 2.0--8.0~keV, 2.0--3.3~keV, 3.3--4.7~keV, and 4.7--8.0~keV
bands. We estimated the photon flux from each source, 
by dividing the net counts by the average of the effective
area function in each band. Table~\ref{tab:phot} lists the 0.5--2.0 keV 
and the 2--8 keV flux, the latter of which is the sum of the fluxes 
in the three sub-bands. Figure~\ref{fig:lumdist} displays histograms
of the net counts and fluxes in the 0.5--2.0~keV and 2--8~keV bands.
Histograms of upper limits are also plotted, for sources that were 
detected in one band but not the other.

\begin{figure}
\centerline{\epsfig{file=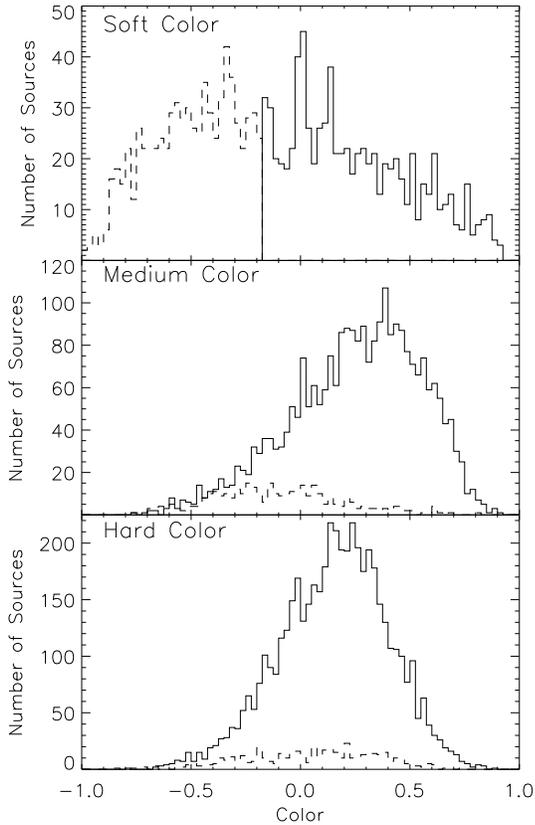,width=0.9\linewidth}}
\caption{Distribution of measured hardness ratios,
$(h-s)/(h+s)$, where 
$h$ and $s$ are the numbers of counts in the higher and lower energy bands, 
respectively. The {\it top}
displays $HR0$, constructed from counts 
in the 2.0--3.3~keV and 0.5--2.0~keV bands; the {\it middle} displays
$HR1$, using counts in the 3.3--4.7~keV bands and 2.0--3.3~keV;
the {\it bottom} displays $HR2$, using counts in the 4.7--8.0~keV and 
3.3--4.7~keV bands. Foreground sources are defined as those with 
$HRO$$<$$-0.175$, and are plotted with the dashed line. Galactic center
sources have $HRO$$\ge$$-0.175$, and are plotted with a solid line. 
Most Galactic center sources do not have measured $HR0$, and their 
$HR1$ is skewed to higher values by absorption.
\label{fig:hrdist}
}
\end{figure}

Finally, using custom code that was not part of ACIS Extract, 
we computed 90\% uncertainties on the net counts in each band, through 
a Bayesian analysis of the Poisson statistics, 
with the simplifying assumption that the uncertainty on the background 
is negligible \citep{kbn91}. 
We used the net counts to compute the hardness ratios $(h-s)/(h+s)$, where 
$h$ and $s$ are the numbers of counts in the higher and lower energy bands, 
respectively. The resulting hardness ratios are bounded by $-1$ and $+1$. 
We defined a soft color using counts 
in the 2.0--3.3~keV and 0.5--2.0~keV bands ($HR0$), a medium color using counts
in the 3.3--4.7~keV bands and 2.0--3.3~keV ($HR1$), and a hard 
color using counts in the 4.7--8.0~keV and 3.3--4.7~keV bands
($HR2$). We calculated uncertainties on
the ratios using the 90\% uncertainties on the net counts and 
Equation~1.31 in Lyons (1991; page 26). 
The hardness 
ratios are listed in Table~\ref{tab:phot}, and histograms showing
their distributions are displayed in Figure~\ref{fig:hrdist}. 

The soft color, $HR0$, was used to distinguish foreground sources from
objects that were likely to lie near or beyond the Galactic center.
We select foreground X-ray sources as those with
soft colors in the range $-1.0$$\le$$HR0$$<$$-0.175$, which corresponds
to absorption columns equivalent to 
$N_{\rm H}$$\la$$4\times10^{22}$~cm$^{-2}$. Most of these
should lie within 4 kpc of Earth \citep[e.g.,][]{mar06}. 
We selected X-ray sources that were 
located near or beyond the Galactic center as those either that had 
soft colors $HR0$$\ge$$-0.175$, 
or that were not detected in either of the 0.5--2.0 and 2.0--3.3~keV bands.
This X-ray selection corresponds to absorption columns equivalent to
$N_{\rm H}$$\ga$$4\times10^{22}$~cm$^{-2}$. 
We find 2257 foreground X-ray sources,
and 6760 sources near or beyond the Galactic center.
Foreground and absorbed sources are plotted separately in
Figure~\ref{fig:hrdist}. Most absorbed sources do not have measured
soft colors. 
 
\subsection{Variability}

We searched for variability using the arrival times of the events. 
We searched for three kinds of variations: 
long-term variations that occurred between observations, 
short-term variability within individual observations, 
and periodic variability within individual observations. 

\subsubsection{Long-term Variability}

We searched for variations that occurred between observations by comparing 
the event arrival times from all of the observations to a constant flux model
using the KS statistic. Any source with a $<$0.1\% chance of being described
by a constant flux model was considered to vary on long time scales. 
There were 856 sources that exhibited long-term variability, 137 of which also 
exhibited short-term variability. Therefore, about 10\% of sources vary 
on the day to month time scales between observations. 

\begin{figure}
\centerline{\epsfig{file=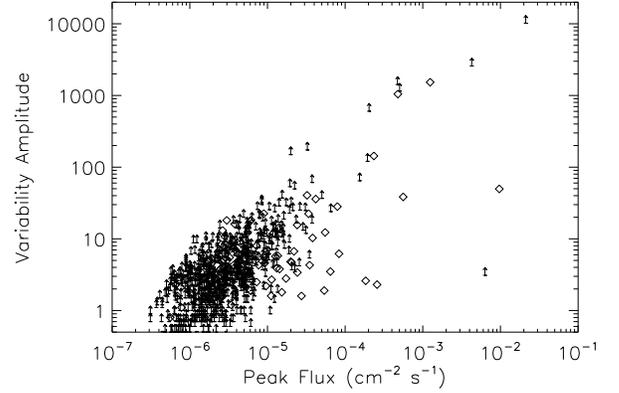,width=0.95\linewidth}}
\caption{
Summary of the properties of long-term variables. We plot the ratio of 
the maximum to minimum fluxes against the maximum flux. 
Measurements are represented with diamonds, and lower 
limits with upward-pointing arrows. The largest-amplitude variations 
necessarily have the largest peak fluxes, because the minimum fluxes
generally represent non-detections, and are therefore equivalent to 
the sensitivity of our observations. 
}
\label{fig:longterm}
\end{figure}

We characterized these long-term variations by computing the mean photon 
flux during each observation of a variable source. Table~\ref{tab:longterm}
lists the source name, observations in which the largest and smallest fluxes 
were observed, the values of the largest and smallest fluxes, and the ratios
of those values. Figure~\ref{fig:longterm}
compares the amplitude of the variations to the maximum flux. 
In order to exclude measurements with poor signal-to-noise, the largest 
flux was defined as the measurement with the largest lower 
limit, and the smallest flux was defined as the measurement with the 
smallest upper limit. In most (740) cases, the smallest flux was consistent 
with zero, and the lower limit to the flux ratios was provided. In 224 cases, 
the uncertainties in the largest and smallest fluxes overlapped, and the
formal lower limit to the ratio was less than 1. The statistics on faint 
sources with low-amplitude variability tended to be poor, so 
Table~\ref{tab:longterm} would be best used to identify highly-variable 
sources for further study. 

\subsubsection{Short-term Variability}

We searched for variability within each observation by comparing the light 
curves to constant count rate models using the KS
statistic. If the 
arrival times of events had a $<$0.1\% chance of being described by a 
uniform distribution, we considered a source to have short-term
variability. We identified 294 sources, or 3\% of our sample, as having 
clear short-term variations.

We roughly characterized the nature of the variability by dividing 
each time series into intervals that were consistent
with having constant count rates, using the Bayesian Blocks algorithm 
of \citet{sca98}. In brief, the algorithm compared the probability that
an interval could be described by two different count rates to the 
null hypothesis that the photons arrived with a single rate. If the 
ratio of the two probabilities exceeded a user-specified prior odds ratio,
then the interval was divided at the time that produced two intervals with
the largest calculated likelihood. This process was iterated until no 
sub-intervals were divided any further. We chose to apply the algorithm
in order to describe each variable light curve with the fewest intervals
with distinct rates (blocks). We applied three progressively 
looser odds ratios, successively demanding that the probability for the 
two-rate model exceed the null hypothesis by factors of 1000, 100, and 10
if the larger odds ratio failed to identify a change point. In this 
way, large flares were described with a few ``blocks'' using a large 
odds ratio, whereas small-amplitude variations were still characterized using
a smaller odds ratio. This approach was deemed necessary in part because 
the Bayesian interpretation of the odds ratios does not have a good 
frequentist analog that could be compared to the probabilities returned by
the KS test, and in part because the KS test and the Bayesian Blocks tests
are most sensitive to slightly different forms of variability. Ultimately,
only 60\% of the variable sources identified with the KS test were 
characterized with more than one block in the Bayesian Block algorithm.

\begin{figure}
\centerline{\epsfig{file=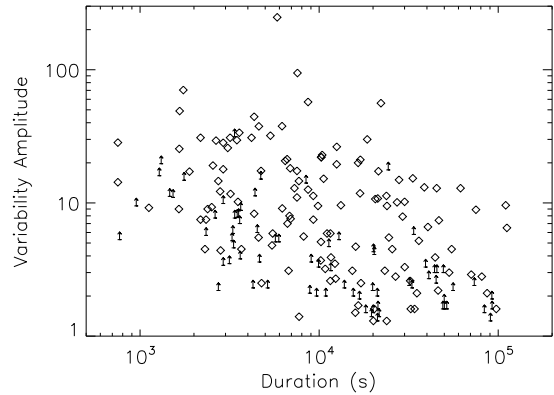,width=0.95\linewidth}}
\caption{
Summary of the properties of short-term variables, using parameters
returned from the Bayesian Blocks algorithm. We plot the ratio of 
the maximum to minimum fluxes against the duration of the peak-flux interval
in an observation. Measurements are represented with diamonds, and lower 
limits with upward-pointing arrows. 
Low-amplitude variations are not represented among 
the short-duration events, because poor counting statistics prevents
us from identifying them. 
}
\label{fig:shortterm}
\end{figure}

Despite the mismatch between the two tests, the characteristics of the
variable sources identified by both the KS and Bayesian Blocks tests
are illustrative. In Table~\ref{tab:shortterm}, we list some properties
of the variable sources: their names, the ObsIDs in which variability 
occurred, the odds ratio at which the Bayesian blocks algorithm identified
a source as variable, the number of blocks used to describe the events, the
durations of the brightest portions of the light curves, the minimum
and maximum fluxes, and the ratios of the maximum to minimum fluxes. 
Figure~\ref{fig:shortterm} compares the 
duration and amplitude of the variability. 
For 40\% of the variable sources, the minimum flux was consistent 
with zero, so the ratio represents a lower limit to the variability amplitude.
We find that all variations had time scales of $>$10 minutes. The
amplitudes ranged from barely-detectable 30\% variations in the flux 
(CXOUGC J174534.8--290851), to one flare in which the flux increased by 
a factor of 250 (CXOUGC J174700.7--283205). Foreground sources are 
over-represented among variable sources --- they compose only 25\% of 
our entire catalog, but 50\% of the short-term variables --- which 
is consistent with the expectation that they are nearby K and M dwarf 
flare stars \citep[e.g.,][]{lay05}. 

\subsubsection{Periodic Variability}

We searched for periodic variability in the brightest sources by 
adjusting the arrival times of their photons to the Solar System barycenter 
and computing Fourier periodograms using the Rayleigh statistic 
\citep{buc83}. 
The individual X-ray events were recorded with a time resolution of 3.2~s, so 
the Nyquist frequency was $\approx$0.15~Hz, which represents the limit 
above which our sensitivity could not be well-characterized. 
However, we computed the periodogram using a maximum frequency of 
$\approx$0.2 Hz, to take advantage of the limited sensitivity to higher 
frequency signals, and to ensure that any observed signal was not an alias. 

We considered sources that, in individual
observations, produced a large enough number of counts ($N_{\gamma}$) 
that a fully-modulated signal could 
be detected with 99\% confidence. The power $P_{\rm meas}$
required to ensure that a source had a chance $<$$1-C$ of being produced 
by white noise can be computed if one knows the number of trials in a search,
($N_{\rm trial}$), and is given by inverting 
$C \approx N_{\rm trial} e^{-P_{\rm meas}}.$
Here, $P_{\rm meas}$ is normalized to have a mean value of 1, and 
the approximation is valid for $P_{\rm meas}$$\gg$1 \citep{rem02}.
A count threshold can be determined by noting that, if background photons
are negligible, the fractional root-mean-squared amplitude of a sinusoidal 
signal ($A$) is given by $A \approx (2P_{\rm meas}/N_\gamma)^{1/2}$. 
A fully-modulated
signal has $A$=0.71. After iterating to determine the number of 
trials corresponding to each count limit, we found that a source
with $N_\gamma$=86 could be identified with $C$=0.99 
if it produced a fully-modulated signal.

In total, we searched for pulsations in 717 event lists from 256 different 
sources, which required $2\times10^7$ trials. A single signal that had 
$C$$>$0.99 given this number of trials must have had $P_{\rm meas}$$>$21.4.
However, multiple observations were searched for many sources, so we 
recorded signals with lower powers and checked whether they also appeared
at the same frequency in other observations.

\begin{figure}
\centerline{\epsfig{file=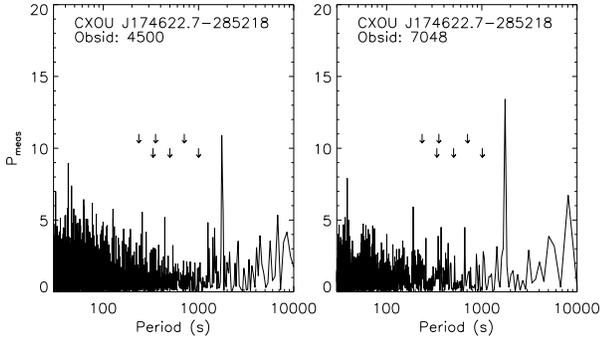,width=0.95\linewidth}}
\caption{
Fourier periodograms for the two observations in which a 1745~s signal
was detected from CXOUGC J174622.7--285218. The signal has a joint probability
of 1.4\% of resulting from white noise, given $N_{\rm trial}$=$2\times10^7$.
The downward-pointing arrows show the fundamental and first two 
harmonics of the periods with which the satellite point was dithered in 
pitch and yaw. 
}
\label{fig:fft}
\end{figure}

We identified two sources with periodic variability at $>$99\%
confidence, CXOUGC J174532.7--290550
and CXOUGC J174543.4--285841. We had previously identified both of these
by combining 500~ks of exposure over the course of two weeks \citep{m-per}. 
The other sources in \citet{m-per} were too faint for their periodic 
variability to be identified in individual observations. We also identified 
a third source as a good candidate for having periodic variability, 
CXOUGC J174622.7--285218. Signals from this source were identified with 
periods of $\approx$1745~s in observation 4500
with $P_{\rm meas}$=10.9 from 763 photons, and 
in observation 7048 with $P_{\rm meas}$=13.4 from 310 photons 
(Figure~\ref{fig:fft}). 
The joint probability that these signals were produced at the same 
frequency by noise \citep{rem02}, 
given $N_{\rm trial}$=$2\times10^7$, was only 1.4\%. 
Periodic signals were not detected from this source in observation
945 because the source fell on a chip edge, nor in observations 2273 and 2276 
because their exposures were too short.

\begin{figure}
\centerline{\epsfig{file=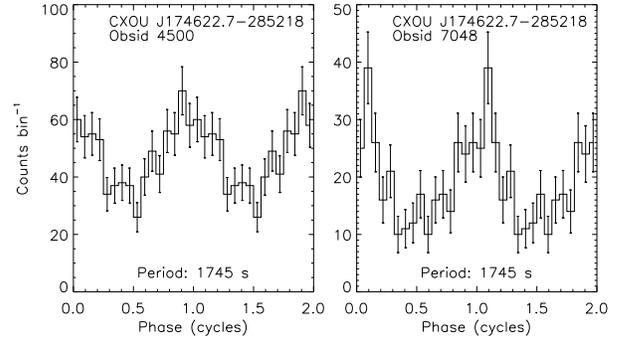,width=0.95\linewidth}}
\caption{
Pulse profiles for the two observations in which the 1745~s signal
was detected from CXOUGC J174622.7--285218. Two identical cycles have been are
displayed in each panel. The profiles are consistent with sinusoids, within
their uncertainties.
}
\label{fig:prof}
\end{figure}

We refined our initial estimates of the period for CXOUGC J174622.7--285218 
for each observation 
by computing pulse profiles 
from non-overlapping $10^4$ s intervals, and modeling
the differences between the assumed and measured phases using a first-order
polynomial. The reference epochs of the pulse maxima for the two 
observations were 53165.3781(6) and 54145.1644(7) (MJD, Barycentric Dynamical
Time).
The best-fit periods were 1745$\pm$3 s and 
1734$\pm$16 s for observations 4500 and 7048, respectively. 
The pulse profiles for each observation are displayed in Figure~\ref{fig:prof}.
The fractional root-mean-squared amplitudes of the pulsations were
21\% and 32\%, respectively. 
Given the long period for this source, it is most likely a 
magnetically-accreting white dwarf \citep{m-per}.

\begin{figure*}[htp]
\centerline{\epsfig{file=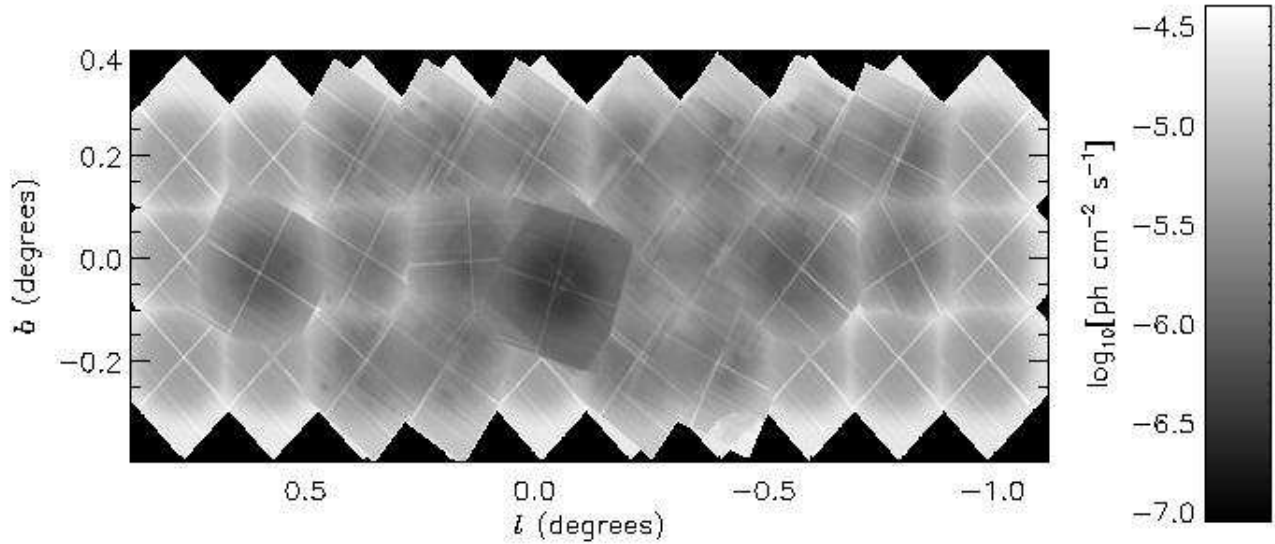,width=0.95\linewidth}}
\caption{
Map of the limiting flux for our survey. Sources brighter than the 
limiting flux at each point have a $>$90\% chance of being detected.}
\label{fig:sens}
\end{figure*}

\begin{figure}
\centerline{\epsfig{file=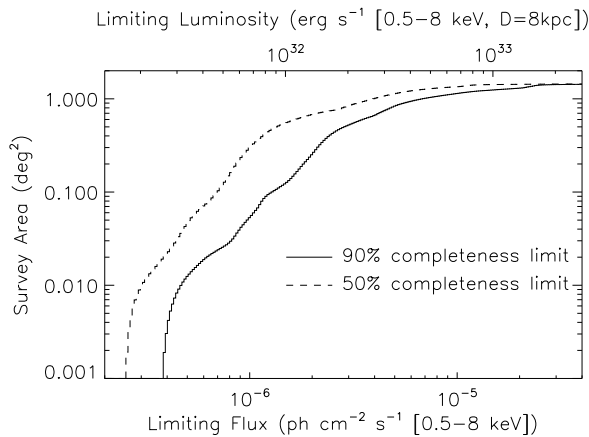,width=0.95\linewidth}}
\caption{
The area over which we were sensitive to sources of given fluxes ({\it
bottom axis}) and luminosities ({\it top axis}, assuming $D$=8~kpc,
a $\Gamma$=0.5 power-law spectrum, and 
$N_{\rm H}$=$6\times10^{22}$~cm$^{-2}$),
with 50\% and 90\% confidence. 
}
\label{fig:area}
\end{figure}

\subsection{Sensitivity}

We calculated the sensitivity of our observations using synthetic-star 
tests, following the basic methods described in 
\citet{bau04} and Muno \etal\ (2006a; see Wang 2004 for another approach). 
We generated maps of our sensitivity 
both for each of the stacked observations 
(i.e., centered on ObsIDs 3392, 4500, 5892, 7034--7048, and 944),
and for a fiducial field with an exposure time of 12 ks for those
regions only covered by the shallow exposures of \citet{wgl02}.
In brief, for each pointing,
we generated a background map by (1) removing events from within a circle 
circumscribing $\approx$90\% of the energy of the PSF around each detected 
source, and then (2) filling the ``holes'' in the image with  
numbers of counts drawn from Poisson distributions with means equal to
those of surrounding annuli. We then simulated 100 star fields per pointing.
We placed $\approx$5000 point
sources at random positions in each background image, with fluxes distributed 
as $N(>S) \propto S^{-\alpha}$ with a slope 
$\alpha = 1.5$, and minimum fluxes that would produce 3 counts in a
100~ks exposure. 
We converted these fluxes to expected values for the numbers of counts 
using an exposure map. The exposure map was normalized to produce the mean 
flux-to-counts conversion for X-ray
sources located at or beyond the Galactic center 
\citep[$HR0 > -0.175$;][]{m-wide}.\footnote{We note that in \citet{m-wide},
we calculated flux limits from a mono-energetic exposure map that
over-estimated the effective area by 50\%, which caused us to report limiting
fluxes that were erroneously low.}
Then, to account for the Eddington bias, we 
drew observed numbers of counts from Poisson distributions with 
mean values equal to the expected counts. 
Next, we obtained model images of the PSF from the routine {\tt mkpsf},
averaged them when appropriate, and used the composite PSF as the 
probability distribution to simulate the 2-dimensional image of 
the counts. These were added to the synthetic exposure. 
Finally, we searched the synthetic image for point sources 
using \program{wavdetect} for the 12 ks exposures, and {\tt wvdecomp} 
for the stacked observations. By comparing the input and output lists, we 
estimated the minimum flux at which a source would be detected in 50\% and
90\% of trials over a grid of points covering our survey. We interpolated 
between these points to make a map of the sensitivity for each image.

None of our observations are formally confusion-limited 
at our completeness limits (Hogg 2001; see also Muno \etal\ 2003). 
If the background diffuse X-ray emission
is unresolved stellar sources, then confusion caused by undetected sources
is accounted for naturally by our background maps.

In order to produce a global sensitivity map, we combined the sensitivity 
maps from the above simulations by recording the best sensitivity at 
each point in the image. The map of 90\% confidence 
limits is displayed in Figure~\ref{fig:sens}. The
effective area of the survey as a function of limiting photon
flux and luminosity is displayed in Figure~\ref{fig:area}. 
We are sensitive to 
$\approx$$4\times10^{32}$~\ergsec\ (0.5-8.0 keV, assuming $D$=8~kpc) 
at 90\% confidence over one 
square degree, and to $\approx$$1\times10^{32}$~\ergsec\ over
0.1 square degrees. This is a factor of $\approx$2 improvement over
\citet{m-wide}.

However, we still find that the majority of X-ray sources are
detected at fluxes below our completeness limits. Of 6760
sources that are likely to lie at or beyond the Galactic center
($HR0 > -0.175$), only 15\% are brighter than the 90\% completeness
limit at the point at which they were detected, and only 40\% are 
brighter than the 50\% completeness limit. This is caused by two 
effects. First, 20\% of the sources are detected only in the hard 
band, whereas our completeness limits are for the full band. Second, 
the number-flux distribution is steep (\S3.3), such that many of
the faint sources are only detected because of positive Poisson fluctuations
in their count rates. 

These maps are used in selecting complete samples of sources for measuring
the spatial (\S3.2) and flux (\S3.3) distributions. Sources below our 
completeness limits are still securely detected, although other sources 
with similar intrinsic fluxes have been missed. 

\subsection{Refined Source Positions}

Experience with matching X-ray and optical sources as part of the
\chandra\ Deep Fields and Orion Ultradeep projects 
suggests that the positions of the
X-ray sources can be refined with respect to those provided by
the wavelet algorithms \citep{alex03,get05}. Therefore, we 
used the implementations of their techniques in ACIS Extract
to refine the positions of our sources.
For each source, we made a composite 
image by combining the event lists from each relevant observation,
and then made a matching composite PSF image that we weighted by the 
values of the exposure maps at the source positions. From this image, 
we computed two additional estimates for the source position:
the mean positions of the events within each source region, 
and a centroid determined by cross-correlating the 
PSF and source images. Following \citet{get05}, if a source lay
within 5\arcmin\ of the aim point, we used
the mean position of the events within the source extraction region. 
If the source lay beyond 5\arcmin, we used the position determined by 
cross-correlating the source image and PSF image. However, if the offset 
of the refined position from the wavelet
position was larger than the smallest source extraction radius that we
used, we assumed that a nearby source had caused confusion, and 
retained the wavelet position. 

Unfortunately, we could not empirically calibrate the uncertainties on 
our source positions, because even the foreground infrared sources 
had such a high density that $\approx$50\% of those that
fell within 3\arcsec\ of an X-ray source were chance alignments. Therefore, 
we computed 95\% positional uncertainties using Equation~5 in 
\citet{hon05},\footnote{This differs from the equation we used for 
eliminating duplicates, because that step was implemented much earlier 
in the process of producing the catalog, before we had settled on a final
uncertainty estimate. The difference has no practical impact on the 
catalog.}
which is based on the positions of sources reported by 
{\tt wavdetect} in simulated observations:
\begin{eqnarray}
r_{\rm err} & = & 0\farcs25 + \frac{0\farcs1}{\log_{10}(c + 1)} \left[1 + \frac{1}{\log_{10}(c + 1)}\right] \nonumber \\
 & + & 0\farcs03 \left[ \frac{\theta}{\log_{10}(c + 2)}\right]^2 +
 0\farcs0006 \left[\frac{\theta}{\log_{10}(c + 3)}\right]^4
 \label{eq:posunc}.  
\end{eqnarray} 
Here, $\theta$ is the offset in
arcminutes of the source from the nominal aim point, and $c$ is the
net number of counts. For sources detected in composite images, we
defined $c$ to be the net counts summed over all observations, 
and $\theta$ to be the
exposure-weighted averages of the sources' offsets from the aim
points of their respective observations. For sources detected in
individual observations, we defined $c$ and $\theta$ to be the values
for the observations in which the sources were identified. 

If $r_{\rm
err}$ is larger than the smallest radius of the region used to extract
photometry for the source (the ``source radius''), the uncertainty 
was set equal to the radius of the extraction region. These sources 
are marginally detected, and the high background in the Galactic center
produces a large tail in the distribution of possible positions. We 
retained them because they passed all of our other selection criteria.

\citet{hon05} established equation~\ref{eq:posunc} by
running {\tt wavdetect} on simulated, single observations that were
generated using a ray-tracing code. Unfortunately, our observations
are more complicated. On the one hand, most of the positions are
determined from composite images generated from observations with very
different aim points. The inclusion of data with large $\theta$ 
could add uncertainty to our measurements. On the other hand,
our positions have been refined compared to the {\tt wavdetect}
values, so the uncertainty on some sources could be smaller.
Therefore, we view Equation~\ref{eq:posunc} as a compromise. Nonetheless,
a comparison of the offsets between 500 foreground X-ray sources and 
the blue 2MASS sources that are their counterparts (as described 
in detail in J. Mauerhan \etal, in prep) reveals that the positions 
in the new catalog are $\approx$60\% better than in \citet{m-cat} 
and \citet{m-wide}.

We also note that because of the way we averaged
the PSF, the positions and uncertainties for the of sources that 
vary in flux between observations (10\% of our sample) could be 
mis-estimated. For example, a variable source that was only bright 
in an off-axis observation would have a larger uncertainty than might 
be expected if it were also bright during an on-axis observation. We 
have not evaluated whether systematic offsets in the positions are 
expected.

\subsection{Details of the Tables\label{sec:tdesc}}

Table~\ref{tab:positions} contains the locations of the point sources, 
parameters related to the observations of each source, and information
on the data quality. Its columns are as follows:

\noindent
(1) Record locators that can be used to cross-correlate 
with other tables.

\noindent
(2) The source names, which are derived from the coordinates of the
source based on the IAU format, in which least-significant figures are
{\it truncated} (as opposed to rounded). The names should not be used 
as the locations of the sources.

\noindent
(3--4) The right ascensions and declinations of the sources, in degrees
(J2000).

\noindent
(5) The 95\% uncertainties in the positions (the error circles).
There are 5810 sources with uncertainties $\le$1\arcsec\ (half of 
which are within 7\arcmin\ of \sgrastar), and 1950 with uncertainties 
$\le$0\farcs5 (85\% of which are within 7\arcmin\ of \sgrastar).

\noindent
(6) Flag indicating how the positions were derived. A ``d''
indicates the position is from the mean position of events, a ``c''
indicates it was derived by cross-correlating the image and the PSF,
and a ``w'' indicates it was derived from a wavelet algorithm. Sources
marked with a ``w'' are likely to be confused with a nearby source, or 
in a region of high background.

\noindent
(7) The images in which the sources were identified. The tags
``full'', ``2002'', ``2004'', and ``2005'' indicate a source
was found in composite images of the \sgrastar\ field. All other values
are the observations in which a source was detected. Two sources added
manually are tagged with ``hand''.

\noindent
(8) Additional information about how the sources were detected.
The tag ``full'' refers to any source detected with {\tt wavdetect}
in the 0.5--8.0~keV 
band; ``soft'' sources were detected in the 0.5--2.0~keV but not the 
full band; ``hard'' sources were detected in the 4--8~keV band but 
neither of the other two bands. The tag ``tile'' indicates that the 
source was detected in a composite 0.5--8.0~keV image with {\tt wvdecomp}.

\noindent
(9) The offsets ($\theta$) from the aim point, in arcmin. If a source 
position was estimated from a composite image, $\theta$ is the mean 
offset weighted by the exposure. If a position was taken from a single
observation, $\theta$ is the offset for that observation. 

\noindent
(10) The number of observations used to compute the photometry 
for each source.

\noindent 
(11) The exposure times in seconds.

\noindent
(12) The fractions of the PSF enclosed by the source extraction
regions.

\noindent
(13) The fiducial energies of the PSFs used to construct the source
extraction regions.

\noindent 
(14) The smallest radius for the extraction region that was
used for a source, in arcseconds. This is determined from the
observation in which the source was closest to the aim point. It is 
also an absolute upper bound to the positional uncertainty for a
source.

\noindent
(15) The 50\% completeness limit at the position of the source. Sources
brighter than these completeness limits can be used to compute spatial
and flux distributions, although the sensitivity map (Fig.~\ref{fig:sens})
is needed to compute the corresponding survey area. 

\noindent
(16) Flags denoting quality, and other information: 
``a'' for sources used to register the astrometry of fields;
``s'' for sources variable on short time scales, 
as indicated by probabilities of $<$0.1\% that the
event arrival times for at least one observation were consistent with
a uniform distribution according to the KS test; ``l'' for sources that were
variable on long time scales, as indicated by a probability of $<$0.1\% 
that the fluxes for all observations were consistent with a uniform 
distribution according to the KS test; ``e'' for sources that may be 
part of an extended, diffuse feature \citep{m-diff};
``c'' for sources confused with another nearby source;
``g'' for sources that fell near the edge of a detector in one or more 
observations; ``b'' for sources for which the source and background 
spectra have a $>$10\% chance of being drawn from the same distribution 
according to a KS test; ``x'' for sources
for which the 0.5--2.0~keV band photometry is inaccurate because the 
satellite was programmed to omit photons below 1~keV from the telemetry;
and ``p'' for sources that suffered from photon pile-up.

Table~\ref{tab:phot} contains the X-ray photometry for each source. 
It contains the following columns

\noindent
(1) The record locators.

\noindent
(2) The source names.

\noindent
(3) The log of the probabilities that the source and background spectra are
derived from the same distribution, according to a
KS test. Large negative values indicate
that the source and background spectra are distinct, and therefore
that the source is most likely real.

\noindent
(4) The total numbers of counts in the 0.5--2.0 keV band.

\noindent
(5) The estimated numbers of background counts in the 0.5--2.0 keV 
band. 

\noindent
(6) The net numbers of counts in the 0.5--2.0 keV band,
and the 90\% lower and upper uncertainties.
In the case of non-detections,
an upper limit is provided.

\noindent
(7) The total numbers of counts in the 2--8 keV band.

\noindent
(8) The estimated numbers of background counts in the 2--8 keV 
band. 

\noindent
(9) The net numbers of counts in the 2--8 keV band, and
the 90\% lower and upper uncertainties. In the case of non-detections,
an upper limit is provided.

\noindent
(10) The fluxes in the 0.5--2.0 keV band, in units of \phcms.

\noindent
(11) The fluxes in the 2--8 keV band, in units of \phcms.

\noindent
(12) The mean energy of photons in the source region, statistically 
corrected for the background. 

\noindent
(13) The soft colors and 90\% upper and lower uncertainties.

\noindent
(14) The medium colors and 90\% upper and lower uncertainties.

\noindent
(15) The hard colors and 90\% upper and lower uncertainties.

These tables were designed to be inclusive, so sources of questionable
quality are included. For instance, 134 sources have net numbers of
counts in the 0.5--8.0 keV band that are consistent with 0 at
the 90\% confidence level. These
sources are only detected in a single band and are presumably either
very hard or very soft, detected in single observations because they were
transients, or detected in stacked observations with {\tt wvdecomp}
at marginal significance. We have chosen to include them 
because they passed the test based on Poisson statistics
from \citet{wei07}.

\begin{figure*}[tbhp]
\centerline{\epsfig{file=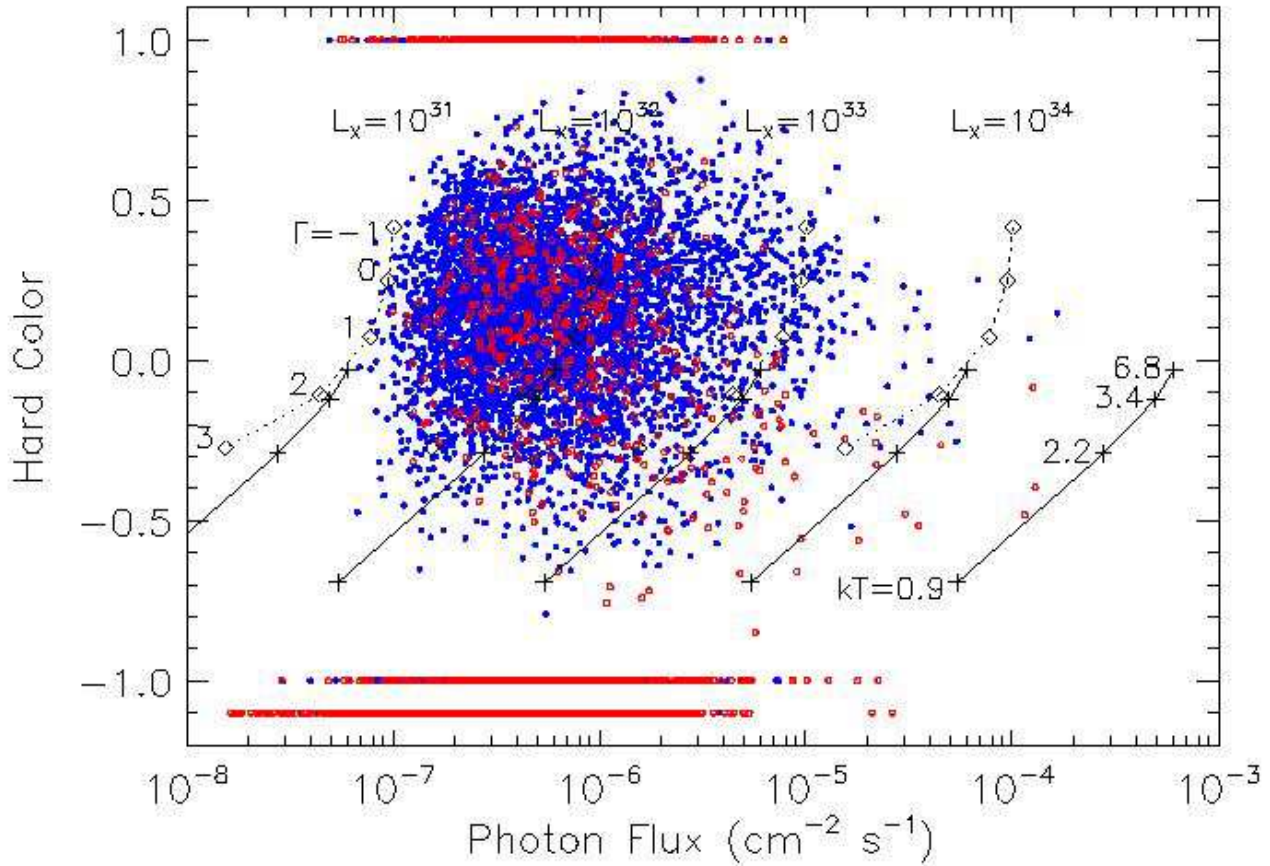,width=0.95\linewidth}}
\caption{The hard color plotted against the photon flux from
each source. Foreground sources are plotted as open red circles, and
Galactic center sources as filled blue circles. 
Sources detected in only in the 3.3--4.7 keV band are assigned hard
colors of $-$1; those only detected in the 4.7--8.0 keV band are assigned
$HR2$=$+$1, and those detected in neither band are assigned $HR2$=$-$1.1.
We also have plotted the colors expected for sources of 
varying luminosities at a distance of 8 kpc, and absorbed by 
$6\times10^{22}$ cm$^{-2}$ of interstellar gas and dust. The dotted
lines are for power-law spectra, and the solid lines for thermal plasma
spectra. 
}
\label{fig:hit}
\end{figure*}

\section{Results}

With a catalog of X-ray sources and associated maps of our sensitivity,
it is straightforward to examine the flux and spatial distributions
of our sources. We have previously reported these quantities based on
the catalogs produced for the central 20 pc around \sgrastar\ 
\citep{m-cat} and on the wide, shallow survey data that was in the 
archive as of 2005 June \citep{m-wide}. Here, 
we derive these quantities for the new catalog, and briefly
compare the distributions to recent results from \citet{koy07}
on the distribution of diffuse iron emission.

\subsection{X-ray Colors and Intensity}

In Figure~\ref{fig:hit}, we plot the hard color versus the flux from
each source. Foreground sources are indicated with open red circles, and 
sources at or beyond the Galactic center with filled blue circles.
There are 6381 Galactic center sources and 1091 foreground sources with 
measured hard colors. 
We have calculated the hardness ratios and photon fluxes that we would expect
to get from these energy bands for a variety of spectra and 0.5--8.0 keV
luminosities using \program{PIMMS} and \program{XSPEC}. In 
Figure~\ref{fig:hit}, we
plot the colors and fluxes expected for power-law spectra with the 
dotted lines, and for a optically-thin thermal plasma with the solid lines.
We have assumed a distance of 8 kpc and $6\times10^{22}$ cm$^{-2}$ of
absorption from interstellar gas and dust.

The median hard color for the Galactic center sources is 0.17. For
interstellar absorption, this corresponds to a $\Gamma$$\approx$0.5 power 
law. Using a simulated spectrum, we have determined that 
the photon fluxes can be converted to energy fluxes according to
1~\phcms~$= 8.7\times10^{-9}$~\ergcms\ (0.5--8.0~keV). The de-absorbed 0.5--8.0
keV flux is approximately 1.7 times larger, so that for a distance
$D$=8 kpc, $10^{34}$ \ergsec\ equals $9\times10^{-5}$ \phcms.
The large median 
value of the hard  color is inconsistent with that expected from a 
thermal plasma (of any temperature) attenuated by interstellar gas and dust.
However, our earlier study of the spectra of brighter sources suggest
that intrinsic absorption is present, and that the underlying 
spectrum is consistent with a $kT$=7--9~keV thermal plasma \citep{m-ps}.
For sources that are intrinsically absorbed, the luminosities will be 
significantly higher than implied by Figure~\ref{fig:hit}. 

\begin{figure*}[htp]
\centerline{\epsfig{file=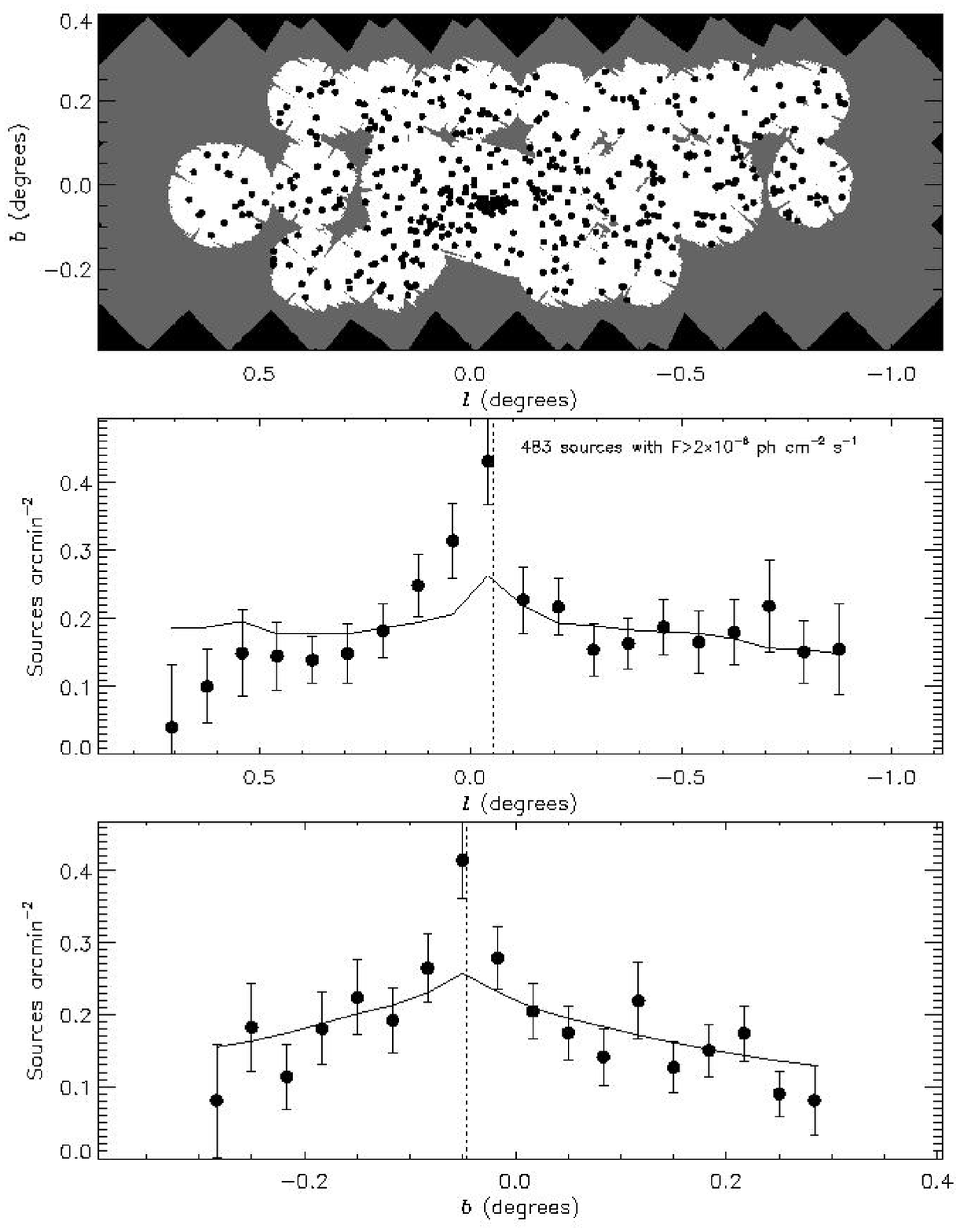,width=0.85\linewidth}}
\caption{
The spatial distribution of point sources that are securely detected,
with fluxes $>$$2\times10^{-6}$ \phcms\ (0.5--8.0 keV), and that 
lie near or beyond the Galactic center ($HR0$$>$$-0.175$). In the 
{\it top panel}, we show the two-dimensional distribution. Over much 
of the region, we are less sensitive than our nominal limit, so 
we have indicated these regions with grey. Regions in which we were 
more sensitive are in white, and the detected sources are indicated
with filled black circles. {\it Middle panel:} Histogram of the 
number of sources per square arcminute, computed as a function of 
Galactic longitude. The area used to normalize the histogram is 
derived from the white area in the panel above.
The solid line illustrates the model stellar distribution from \citet{lzm02}
and \citet{kdf91}, which originally was derived from infrared observations. 
The model distribution also was computed
for the white area in the top panel.
{\it Bottom panel:} Same as for the middle panel, except that the 
source distribution is plotted as a function of Galactic latitude.
}
\label{fig:londist}
\end{figure*}

\subsection{Spatial Distribution}

We present the spatial distribution of X-ray sources located near or 
beyond the Galactic center ($HR0$$>$$-0.175$) in Figure~\ref{fig:londist}. 
We examined only sources brighter than $2\times10^{-6}$ \phcms, and only 
included 
a source if the 50\%-confidence 
flux limit at its position was less than or equal to 
$2\times10^{-6}$ \phcms. This flux limit was chosen as a compromise
between the area over which the distribution is derived, which decreases
for lower flux limits (Fig.~\ref{fig:sens}), and the number of sources
used in the distribution, which tends to increase for lower flux limits
(Fig.~\ref{fig:lumdist}). In the top panel of Figure~\ref{fig:londist}, 
we display the locations of each of the 479 sources that met the flux 
criteria. The area over which the flux limit is $<$$2\times10^{-6}$ \phcms\
is displayed in white, and the greyed areas indicate regions of poorer 
sensitivity. A concentration of X-ray sources is evident near the 
position of \sgrastar. In the bottom panels of Figure~\ref{fig:londist},
we display histograms of the numbers of sources per unit area, as functions
of Galactic longitude and latitude. Only regions of good sensitivity are used.

We then compared the spatial distributions to that of the stellar mass
that has been inferred from infrared observations. Our mass model
consists of the young nuclear bulge and cusp and the old Galactic 
bulge from \citet{lzm02}, and the model for the Galactic disk from 
\citet[][see Muno \etal\ 2006a for further details]{kdf91}. To make a 
direct comparison with our unevenly-sampled spatial distributions, we 
integrated the model for the stellar mass from 6 to 14 kpc along the line of 
sight at points on a 1\arcmin\ grid covering our survey region, and 
interpolated the resulting values onto the image. We then summed the 
values of the 
integrated mass over areas of good sensitivity, to match the longitude and
latitude bins
of the observed histogram. Finally, we minimized chi-squared over one 
parameter to scale the binned 
mass model to the observed distributions of X-ray sources. We find a 
best-fit scaling factor of $5\times10^{-7}$ X-ray sources per solar 
mass for sources brighter than $2\times10^{-6}$ \phcms\ for both the
latitude and longitude distributions. For the longitude distribution, 
$\chi^2/\nu$=22.4/19, and for the latitude distribution
$\chi^2/\nu$=20.6/17.
The best-fit models are displayed with  solid lines in the bottom
panels of Figure~\ref{fig:londist}. 

The models are acceptable descriptions 
of the data. However, in the plot as a function of longitude,
at the inner few arcminutes around \sgrastar, and
just to the east toward the Arches and Quintuplet regions, there is 
an $\approx$2.8$\sigma$ excess in the number of observed X-ray sources. 
We find that this excess is also present with similar significance 
if we chose tighter or looser flux limits between 1 and 
$5\times10^{-6}$~\phcms. The model also predicts more sources than
observed at $L$=0.5--0.6\degree\ at the 1$\sigma$ level, but this is 
probably because the Sgr B molecular complex attenuates X-rays from
sources behind it. 

\subsection{Number-Flux Distribution}

\begin{figure}
\centerline{\epsfig{file=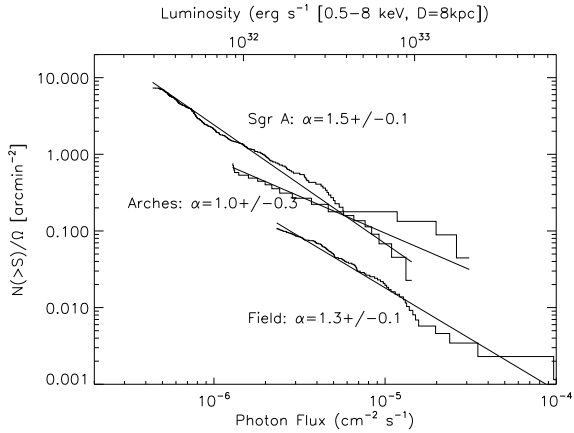,width=0.95\linewidth}}
\caption{
The cumulative number of sources as a function of limiting flux,
for three regions of interest: the inner 8\arcmin\ radius around 
\sgrastar, the 8\arcmin\ radius around the Arches cluster (excluding the 
overlap with the \sgrastar\ region), and the wide survey (excluding
the fields around Sgr B, Sgr C, the Arches, and \sgrastar). 
The solid line indicates the best-fit power law, which we determined
from the un-binned distribution. The top axis provides an estimate of the 
luminosity corresponding to the observed flux. The luminosity is 
calculated assuming $D$=8~kpc and a mean photon energy of 
$8.7\times10^{-9}$ erg (corresponding to a $\Gamma=$=0.5 power law absorbed by 
$N_{\rm H}$=$6\times10^{22}$~cm$^{-2}$).
}
\label{fig:lognlogs}
\end{figure}

We computed the number-flux distribution based on the maximum-likelihood
algorithm described in \citet{mcj73}, which we modified to use 
Poisson statistics in the manner described in Appendix B of \citet{m-wide}. 
We examined three regions that had well-defined flux limits and 
effective exposure times: the inner 8\arcmin\ around \sgrastar,
the 8\arcmin\ around the Arches cluster (excluding the overlap
with the \sgrastar\ field), and the portions of the survey covered
by the 40 ks pointings taken between 2006 and 2007. We assumed that the
number-flux distribution was a single power law over the ranges of 
fluxes that we measured, $N(>S) = N_0 (S/S_0)^{-\alpha}$. 
We display the resulting
cumulative number-flux distributions in Figure~\ref{fig:lognlogs}, and
list the best-fit parameters in Table~\ref{tab:lognlogs}. Our 
distributions extend a factor of $\approx$2 deeper than in 
\citet{m-wide}.

The fit to the distribution from the \sgrastar\ 
region is formally poor, because the distribution steepens at low fluxes
\citep{m-cat}. However, we do find that the Arches region has 
a flatter flux distribution ($\alpha$=1.0$\pm$0.3) than either the
inner 8\arcmin\ around
Sgr A (($1.55$$\pm$0.09) or the wide survey field ($1.3$$\pm$0.1). 
The difference is only significant at the 1.4$\sigma$ level. 
Nonetheless, given that there are also excess sources coincident with the 
Arches region in the spatial distribution, we suggest that there is 
a genuine over-abundance of bright X-ray sources in this region of 
recent star formation. 

\begin{deluxetable*}{lcccccc}[htp]
\tablecolumns{7}
\tablewidth{0pc}
\tablecaption{Parameters of the $\log N - \log S$ Distribution\label{tab:lognlogs}}
\tablehead{
\colhead{Field} & \colhead{$S_{\rm lim}$} & \colhead{Num.} & \colhead{Area} 
& \colhead{$\alpha$} & \colhead{$N_0$} & \colhead{$P_{\rm KS}$} \\
\colhead{} & \colhead{$10^{-6}$ \phcms} & \colhead{Sources} & 
\colhead{(arcmin$^{2}$)} & \colhead{} & \colhead{(arcmin$^{-2}$)} &
\colhead{}
} 
\startdata
Sgr A* & 0.5 & 323 & 44 & 1.55$\pm$0.09 & 0.41 & 0.00 \\
Arches & 1 & 17 & 22 & 1.0$\pm$0.3 & 0.08 & 0.88 \\
Field & 3 & 92 & 813 & 1.3$\pm$0.1 & 0.02 & 0.88 
\enddata
\tablecomments{The normalization of the $\log N - \log S$ distribution,
$N_0$ is listed for a fiducial flux of $2\times10^{-6}$ \phcms, to
match the spatial distribution in Figure~4. $P_{\rm KS}$ represents
the probability under a Kolgoromov-Smirnov
test of seeing the observed difference between the observed 
and model distribution assuming that they are identical, 
so that very small values would indicate a poorer match.}
\end{deluxetable*}

A similar asymmetry has been identified in the flux of diffuse
emission from helium-like iron \citep{koy07}. We suggest that both the
excess point sources and the excess iron emission are related to the
concentration of young stars in this region, the most dramatic
manifestations of which are the Arches and Quintuplet clusters
\citep[e.g.,][]{fig99}. The iron emission is probably diffuse, hot
plasma that forms in shocks where the stellar winds from the clusters
impact the ISM \citep{yz02,lyz04,wdl06}. The excess point sources are probably
young, OB and Wolf-Rayet stars in binaries \citep[e.g.,][]{mau07}.

\section{Discussion}

We have presented a catalog of 9017 X-ray sources located in the inner
2\degree\ by 0.8\degree\ around the Galactic center. This increases
the number of sources known in the region by a factor of 2.5. For all
of the sources, we provide tables listing their positions
(Table~\ref{tab:positions}), photometry, and colors
(Table~\ref{tab:phot}). Of these sources, 6760 have hard colors that
are consistent with high absorptions columns 
$N_{\rm H}$$\ga$$4\times10^{22}$~cm$^{-2}$, which indicates that they lie at
or beyond the Galactic center. 
In addition, the positions of the X-ray sources in this catalog are 
more accurate than earlier versions. This catalog contains 2029 sources with 
$<$0.5\arcsec\ uncertainties (90\% confidence), and another 3981  
with uncertainties between 0.5\arcsec\ and 1\arcsec.
This catalog will be excellent for comparisons
with multi-wavelength ones, in order to search for young stars,
high-mass X-ray binaries, and pulsars 
\citep[e.g.,][]{wll02b,lwl03,mik06,m-ys,mau07}.

The luminosity range that we cover, from $10^{31}$ to 
$10^{34}$~\ergsec\ (0.5--8.0 keV; assuming a $\Gamma$=1.5 power law, 
$N_{\rm H}$=$6\times10^{22}$ cm$^{-2}$, and $D$=8 kpc), is at least
an order of magnitude fainter than studies of Local Group galaxies
\citep[e.g.,][]{tp04,kil05,plu08}. Consequently, the natures of the 
sources that we study are also very different. Whereas the detectable 
stellar population of external galaxies in X-rays is dominated by 
accreting black holes and neutron stars, most of our sources are 
probably cataclysmic variables \citep[e.g.,][]{m-wide}. The hardness of 
the X-ray colors (Fig.~\ref{fig:hit}) suggests that the sources
are specifically magnetically-accreting white dwarfs \citep{ei99,m-wide}.
Therefore, the X-ray population probably represents old stars. 
Indeed, the spatial distribution of sources brighter than 
$2\times10^{-6}$~\phcms\ (2--8 keV) traces that of the old stellar population
(Fig.~\ref{fig:londist}). This makes the population of X-ray sources
in the Galactic center similar to those seen in 
globular clusters \citep[e.g.,][]{ver97,hei06}.
%, although the stellar dynamics in the Galactic
%center should be significantly different \citep[e.g.][]{m-trans}.

Although the distribution of the majority of the X-ray sources
traces that of the old stellar population, we have 
found 2.8$\sigma$ evidence for an excess of sources in two regions
where young, massive stars are forming: in the inner few arcminutes around 
\sgrastar, and in the region where the Arches and Quintuplet star
clusters lie. The excess of sources near these 
young star clusters also appears in the number of sources as a function
of limiting flux, in which relatively more bright X-ray sources are 
found near the Arches and Quintuplet (Fig.~\ref{fig:lognlogs} and 
Table~\ref{tab:lognlogs}). In total, these two regions contain a couple dozen 
more bright sources than our stellar mass model predicts. 
We suggest that these excess X-ray sources
are part of the young stellar population in these region 
\citep{mik06,m-ys,mau07}. In the near future, we will publish additional 
OB and  Wolf-Rayet stars that have been identified through infrared 
spectroscopy of counterparts to X-ray sources (J. Mauerhan \etal, in prep).

\begin{deluxetable*}{llcccl}[htp]
\tabletypesize{\scriptsize}
\tablecolumns{6}
\tablewidth{0pc}
\tablecaption{Luminous X-ray Binaries Covered by Our 
Observations\label{tab:transients}}
\tablehead{
\colhead{\chandra\ name} & \colhead{Common Name} & \colhead{RA} & 
\colhead{DEC} & \colhead{uncertainty} & \colhead{Reference} \\
\colhead{(CXOUGC J)} & \colhead{} & \multicolumn{2}{c}{(Degrees, J2000)} &
\colhead{(arcsec)} & colhead{}
}
\startdata
174354.8-294441 & 1E 1740.7-2942 & 265.97864 & $-29.74499$ &  0.5 & \citet{sid99} \\
174417.2-293943 & AX J1744.3-2940 & 266.07190 & $-29.66234$ &  0.5 & \citet{sid01} \\
174433.0-284427 & Bursting Pulsar & 266.13788 & $-28.74096$ &  0.5 & \citet{ww02} \\ 
174451.6-292042 & KS 1741-293 & 266.21515 & $-29.34522$ &  0.5 & \citet{int97} \\
174457.4-285021 & XMM J174457-2850.3 & 266.23944 & $-28.83917$ &  0.3 & \citet{sak05} \\ [2pt]
174502.3-285449 & Granat 1741.9-2853 & 266.25983 & $-28.91397$ &  0.4 & \citet{m-grs} \\ 
174535.6-290133 & AX J1745.6-2901 & 266.39853 & $-29.02612$ &  0.4  & \citet{mae96} \\
174535.5-290124 & \nodata & 266.39822 & $-29.02337$ &  0.3  & \citet{m-trans}\\
174537.1-290104 & 1A 1742-289 & 266.40494 & $-29.01796$ &  0.4 & \citet{dav76} \\ 
174538.0-290022 & \nodata & 266.40863 & $-29.00623$ &  0.3  & \citet{m-trans} \\  [2pt]
174540.0-290005 & \nodata & 266.41699 & $-29.00160$ &  0.4 & \citet{m-trans} \\ 
174540.0-290030 & \nodata & 266.41684 & $-29.00859$ &  0.3  & \citet{m-trans} \\ 
174540.9-290014 & \nodata & 266.42078 & $-29.00398$ &  0.4  & \citet{m-trans} \\
174553.9-290346 & SWIFT J174553.9-290347 & 266.47467 & $-29.06305$ &  0.4 & \nodata \\
174554.4-285455 & XMM J174554.4-285456 & 266.47690 & $-28.91533$ &  0.4 & \citet{por05} \\ [2pt]
174621.0-284342 & 1E 1743.1-2843 & 266.58768 & $-28.72868$ &  0.4 & \citet{por03} \\
174702.5-285259 & SAX J1747.0-2853 & 266.76080 & $-28.88307$ &  0.4 & \citet{wmw02} \\ 
\nodata & XTE J1748-288 & 267.02108 & --28.47383 & 0.6 & \citet{hrm98} \\
\nodata & XMM J174544-2913.0 & 266.43546 & --29.21683 & 4.0 & \citet{sak05} \\
\enddata
\end{deluxetable*}

A small fraction of the X-ray sources should be accreting black holes
and neutron stars. Around 300 such X-ray binaries are known in the 
Galaxy, about half of which contain low-mass donors that over fill their
Roche lobe, and half of which contain high-mass (OB and Wolf-Rayet) stars
that donate mass through a stellar wind \citep{liu06,liu07}. These 
X-ray binaries are most-easily identified when they are bright 
and variable \citep{m-trans}.
In total, over the history of X-ray astronomy, 19 X-ray sources in 
our survey field have been observed to be $>$$10^{34}$~\ergsec\ in X-rays, 
and have varied by at least an order of magnitude in X-ray flux 
(Table~\ref{tab:transients}). Fifteen of 
these transient X-ray sources were bright during the time span of our
\chandra\ observations (1A 1742-289 and XTE J1748-288 never entered outburst). 
Half of them have been discovered in the last 
9 years using \chandra, \xmm\, or {\it Swift} 
\citep[e.g.,][]{sak05,por05,m-trans,wij06,ken06}. Surprisingly, 
despite having obtained 600 ks of new data in 2006 and 2007, 
we did not detect any new, bright ($>$$10^{34}$~\ergsec), transient 
X-ray sources. This suggests that we have 
identified all of the X-ray binaries that are active on time scales of 
a decade. 

As mentioned in \S\label{ref:obs}, the tables from this work will be
available in the electronic edition of this journal, and additional
products will be made available from the authors' web 
site.\footnote{{\tt http://www.srl.caltech.edu/gc\_project/xray.html}} 
The data available from the authors' site includes FITS images
of all of the images presented in this paper, as well as
the averaged event lists, snapshot images, spectra, and calibration 
files for each source in the catalog. Combined with an increasing amount
of multi-wavelength data, this data set can be used to better understand
the interactions between stars and interstellar media in the Galactic 
center, and the population of X-ray emitting objects in general.

\acknowledgements
MPM, RMB, WNB, GCB, PSB, AC, SDH, JCM, QDW, ZW, and FYZ received
support from NASA thorugh Chandra Award Number G06-7135 issued by the
Chandra X-ray Observatory Center, which is operated by the Smithsonian
Astrophsyical Observatory for and on behalf of the National
Aeronautics Space Administration under contract NAS8-03060. TJWL and
NEK received funding for basic research in astronomy at the Naval
Research Laboratory, which is supported by 6.1 base funding.

\clearpage
\LongTables
\begin{deluxetable*}{lclccccccc}[htp]
\tabletypesize{\scriptsize}
\tablecolumns{10}
\tablewidth{0pc}
\tablecaption{Observations of the Central 2\degree$\times$0.8\degree of the Galaxy\label{tab:obs}}
\tablehead{
\colhead{} & \colhead{} & \colhead{} & 
\multicolumn{2}{c}{Aim Point} & \colhead{} & \multicolumn{3}{c}{Astrometry} \\
\colhead{Start Time} & \colhead{ObsID} & \colhead{Target} & \colhead{Exposure} & 
\colhead{RA} & \colhead{DEC} & \colhead{Roll} & 
$N_{\rm IR}$ & $N_{\rm X}$ & \colhead{Unc.}\\
\colhead{(UT)} & \colhead{} & \colhead{} & \colhead{(ks)} 
& \multicolumn{2}{c}{(degrees J2000)} & \colhead{(degrees)} & 
\colhead{} & \colhead{} & \colhead{(arcsec)}
}
\startdata
2001 Jul 21 04:35:09 & 2286 & GCS 28  & 11.6 & 266.01026 & -29.87409 & 283.8 & \nodata & \nodata & 0.5 \\
2001 Jul 21 08:03:39 & 2289 & GCS 29 & 11.6 & 265.81895 & -29.77175 & 283.8 & \nodata & \nodata & 0.5 \\ 
2001 Jul 21 11:32:10 & 2290 & GCS 30 & 11.6 & 265.62792 & -29.66920 & 283.8 & \nodata & \nodata & 0.5 \\
2000 Aug 30 16:59:32 & 658  & 1E 1740.7-2942 & 9.2 & 265.97607 & -29.75041 & 270.8 & \nodata & \nodata & 0.5 \\
2001 Jul 20 18:09:40 & 2278 & GCS 25 & 11.6 & 266.12796 & -29.70796 & 283.8 & \nodata & \nodata & 0.5 \\ [2pt]
2006 Jun 27 16:38:52 & 7042 & Deep GCS 12 & 14.4 & 265.92331 & -29.62836 & 297.2 & 8 & \nodata & 0.2 \\
2006 Jul 01 13:53:08 & 7346 & Deep GCS 12 & 15.2 & 265.92332 & -29.62846 & 297.2 & \nodata & 8 & 0.3 \\
2006 Jun 28 16:24:41 & 7345 & Deep GCS 12 & 10.0 & 265.92328 & -29.62841 & 297.2 & \nodata & 2 & 0.4 \\
2001 Jul 20 21:38:10 & 2281 & GCS 26 & 11.6 & 265.93676 & -29.60580 & 283.8 & \nodata & 8 & 0.2 \\
2007 Apr 29 06:03:16 & 7043 & Deep GCS 13 & 34.2 & 265.72756 & -29.51198 & 82.9 & 9 & \nodata & 0.13 \\ [2pt]
2001 Jul 21 01:06:39 & 2283 & GCS 27 & 11.6 & 265.74591 & -29.50336 & 283.8 & \nodata & 7 & 0.2 \\
2001 Jul 20 08:00:49 & 2270 & GCS 22 & 10.6 & 266.24516 & -29.54168 & 283.8 & \nodata & \nodata & 0.5 \\
2005 Jul 22 01:58:12 & 5892 & Sgr C & 97.9 & 266.08819 & -29.43665 & 278.0 & 25 & \nodata & 0.08 \\
2001 Jul 20 11:12:40 & 2272 & GCS 23 & 11.6 & 266.05414 & -29.43966 & 283.8 & \nodata & 15 & 0.2 \\
2006 Sep 15 14:54:55 & 7041 & Deep GCS 10 & 19.9 & 265.84657 & -29.35563 & 269.0 & 4 & \nodata & 0.2 \\  [2pt]
2006 Sep 27 01:21:42 & 8214 & Deep GCS 10 & 17.7 & 265.84656 & -29.35560 & 269.0 & 4 & \nodata & 0.2 \\
2001 Jul 20 14:41:10 & 2275 & GCS 24 & 11.6 & 265.86352 & -29.33749 & 283.8 & \nodata & 4 & 0.2 \\
2007 Apr 25 14:04:05 & 7040 & Deep GCS 8 & 36.7 & 266.35283 & -29.38207 & 84.0 & 15 & \nodata & 0.10 \\
2001 Jul 19 06:41:38 & 2296 & GCS 19 & 11.1 & 266.36200 & -29.37542 & 283.8 &  \nodata & 6 & 0.2 \\
2006 Oct 31 03:58:52 & 7038 & Deep GCS 6 & 19.7 & 266.15861 & -29.29070 & 268.2 & 3 & \nodata & 0.2 \\ [2pt]
2007 Feb 19 01:49:07 & 8459 & Deep GCS 6 & 19.3 & 266.15788 & -29.27900 & 91.6 & \nodata & 16 & 0.3 \\
2001 Jul 19 10:01:48 & 2267 & GCS 20 & 8.7 & 266.17124 & -29.27356 & 283.8 & \nodata & 11 & 0.2 \\
2006 Jul 27 05:11:38 & 7039 & Deep GCS 7 & 37.8 & 265.96598 & -29.18723 & 276.1 & 7 & \nodata & 0.17 \\
2001 Jul 20 04:37:11 & 2268 & GCS 21 & 10.8 & 265.98088 & -29.17148 & 283.8 & \nodata & 7 & 0.2 \\
2006 Aug 24 20:26:37 & 7037 & Deep GCS 5 & 39.4 & 266.47130 & -29.22528 & 275.2 & 12 & \nodata & 0.12 \\ [2pt]
2001 Jul 18 20:49:28 & 2291 & GCS 16 & 10.6 & 266.47846 & -29.20889 & 283.8 & \nodata & 11 & 0.2 \\
2006 Jul 26 18:06:34 & 7035 & Deep GCS 3 & 38.0 & 266.27807 & -29.12211 & 276.3 & 5 & \nodata & 0.12 \\
2001 Jul 19 00:01:18 & 2293 & GCS 17 & 11.1 & 266.28790 & -29.10730 & 283.8 & \nodata & 21 & 0.2 \\
2007 Jul 19 20:45:11 & 8567 & Deep GCS 4 & 19.5 & 266.09028 & -29.02150 & 278.3 & 10 & \nodata & 0.18 \\
2007 Jul 09 16:12:13 & 7036 & Deep GCS 4 & 19.9 & 266.09096 & -29.02114 & 283.5 & 7 & \nodata & 0.15 \\  [2pt]
2001 Jul 19 03:21:28 & 2295 & GCS 18 & 11.1 & 266.09774 & -29.00538 & 283.8 & \nodata & 8 & 0.2 \\
2001 Jul 14 01:51:10 & 1561b& \sgrastar\ & 13.5 & 266.41344 & -29.01281 & 264.7 & \nodata & 24 & 0.12 \\ 
2000 Oct 26 18:15:11 & 1561a& \sgrastar\ & 35.7 & 266.41344 & -29.01281 & 264.7 & \nodata & 68 & 0.06 \\
2007 Jul 20 02:27:01 & 7557 & \sgrastar\ & 5.0 & 266.42069 & -29.01498 & 278.4 & \nodata & 9 & 0.15 \\
2004 Aug 28 12:03:59 & 5360 & \sgrastar\ & 5.1 & 266.41477 & -29.01211 & 271.0 & \nodata & 9 & 0.1 \\  [2pt]
2005 Jul 29 19:51:11 & 5952 & \sgrastar\ & 43.1 & 266.41508 & -29.01219 & 275.5 & \nodata & 76 & 0.06 \\
2005 Jul 30 19:38:31 & 5953 & \sgrastar\ & 45.4 & 266.41506 & -29.01218 & 275.3 & \nodata & 78 & 0.06 \\ 
2005 Jul 24 19:58:27 & 5950 & \sgrastar\ & 48.5 & 266.41519 & -29.01222 & 276.7 & \nodata & 64 & 0.06 \\
2005 Jul 27 19:08:16 & 5951 & \sgrastar\ & 44.6 & 266.41512 & -29.01219 & 276.0 & \nodata & 66 & 0.06 \\
2006 Jul 17 03:58:28 & 6363 & \sgrastar\ & 29.8 & 266.41541 & -29.01228 & 279.5 & \nodata & 59 & 0.06 \\ [2pt]
2006 Jul 30 14:30:26 & 6643 & \sgrastar\ & 5.0 & 266.41510 & -29.01218 & 275.4 & \nodata & 11 & 0.18 \\
2005 Aug 01 19:54:13 & 5954 & \sgrastar\ & 18.1 & 266.41502 & -29.01215 & 274.9 & \nodata & 36 & 0.1 \\ 
2006 Sep 25 13:50:35 & 6645 & \sgrastar\ & 5.1 & 266.41448 & -29.01195 & 268.3 & \nodata & 7 & 0.18 \\
2006 Aug 22 05:54:34 & 6644 & \sgrastar\ & 5.0 & 266.41484 & -29.01202 & 271.7 & \nodata & 10 & 0.18 \\ 
2004 Jul 06 22:29:57 & 4684 & \sgrastar\ & 49.5 & 266.41597 & -29.01236 & 285.4 & \nodata & 94 & 0.06 \\ [2pt]
2004 Jul 05 22:33:11 & 4683 & \sgrastar\ & 49.5 & 266.41605 & -29.01238 & 286.2 & \nodata & 98 & 0.06 \\
2006 Oct 29 03:28:20 & 6646 & \sgrastar\ & 5.1 & 266.41425 & -29.01178 & 264.4 & \nodata & 9 & 0.18 \\
2006 Jul 04 11:01:35 & 6642 & \sgrastar\ & 5.1 & 266.41633 & -29.01237 & 288.4 &\nodata & 7 & 0.18 \\
2003 Jun 19 18:28:55 & 3549 & \sgrastar\ & 24.8 & 266.42092 & -29.01052 & 346.8 & \nodata & 49 & 0.06 \\
2006 Jun 01 16:07:52 & 6641 & \sgrastar\ & 5.1 & 266.42018 & -29.00440 & 69.7 & \nodata & 7 & 0.18 \\ [2pt]
2002 May 22 22:59:15 & 2943 & \sgrastar\ & 34.7 & 266.41991 & -29.00407 & 75.5 & \nodata & 86 & 0.15 \\ %-.21 -0.15
2002 Jun 03 01:24:37 & 3665 & \sgrastar\ & 89.9 & 266.41992 & -29.00407 & 75.5 & \nodata & 162 & 0.06 \\
2002 May 25 15:16:03 & 3392 & \sgrastar\ & 165.8 & 266.41992 & -29.00408 & 75.5 & 23 & \nodata & 0.06 \\
2002 May 28 05:34:44 & 3393 & \sgrastar\ & 157.1 & 266.41992 & -29.00407 & 75.5 & \nodata & 759 & 0.06 \\
2002 May 24 11:50:13 & 3663 & \sgrastar\ & 38.0 & 266.41993 & -29.00407 & 75.5 & \nodata & 82 & 0.10 \\  [2pt] 
2006 Apr 11 05:33:20 & 6639 & \sgrastar\ & 4.5 & 266.41890 & -29.00369 & 86.2 & \nodata & 11 & 0.18 \\ 
2006 May 03 22:26:26 & 6640 & \sgrastar\ & 5.1 & 266.41935 & -29.00383 & 82.8 & \nodata & 8 & 0.18 \\
2005 Feb 27 06:26:04 & 6113 & \sgrastar\ & 4.9 & 266.41870 & -29.00353 & 90.6 & \nodata & 6 & 0.12 \\
2002 May 07 09:25:07 & 2954 & \sgrastar\ & 12.5 & 266.41938 & -29.00374 & 82.1 & \nodata & 25 & 0.15 \\
2007 Feb 11 06:16:55 & 7554 & \sgrastar\ & 5.1 & 266.41853 & -29.00343 & 92.6 & \nodata & 11 & 0.2 \\ [2pt]
2002 Feb 19 14:27:32 & 2951 & \sgrastar\ & 12.4 & 266.41867 & -29.00335 & 91.5 & \nodata & 35 & 0.15 \\ % -0.29, 0.03
2002 Mar 23 12:25:04 & 2952 & \sgrastar\ & 11.9 & 266.41897 & -29.00343 & 88.2 & \nodata & 23 & 0.15 \\ %0.25, -0.25
2002 Apr 19 10:39:01 & 2953 & \sgrastar\ & 11.7 & 266.41923 & -29.00349 & 85.2 & \nodata & 28 & 0.15 \\ % 0.07 0.03 
2007 Mar 25 22:56:07 & 7555 & \sgrastar\ & 5.1 & 266.41420 & -29.00013 & 88.0 & \nodata & 10 & 0.15 \\ 
2001 Jul 18 11:13:58 & 2282 & GCS 13 & 10.6 & 266.59457 & -29.04224 & 283.8 & \nodata & \nodata & 0.5 \\ [2pt]
2001 Jul 18 14:25:48 & 2284 & GCS 14 & 10.6 & 266.40415 & -28.94090 & 283.8 & \nodata & 23 & 0.2 \\ 
2007 Feb 22 03:29:33 & 7034 & Deep GCS 1 & 39.6 & 266.20175 & -28.83783 & 91.0 & 36 & \nodata & 0.03 \\
2001 Jul 18 17:37:38 & 2287 & GCS 15 & 10.6 & 266.21412 & -28.83905 & 283.8 & \nodata & 18 & 0.2 \\
2000 Jul 07 19:05:19 & 945 & GC Arc & 48.8 & 266.58221 & -28.87193 & 284.4 & 12 & \nodata & 0.12 \\
2004 Jun 09 08:50:32 & 4500 & Arches Cluster & 98.6 & 266.48260 & -28.81691 & 55.2 & 26 & \nodata & 0.10 \\ [2pt]
2007 Feb 14 03:59:44 & 7048 & Deep GCS 27 & 38.2 & 266.71020 & -28.87828 & 92.3 & 6 & \nodata & 0.17 \\
2001 Jul 18 00:48:28 & 2273 & GCS 10 & 11.2 & 266.71024 & -28.87570 & 283.8 & \nodata & 6 & 0.2 \\
2001 Jul 18 04:16:58 & 2276 & GCS 11 & 11.6 & 266.52002 & -28.77435 & 283.8 & \nodata & 12 & 0.2 \\
2001 Jul 18 07:45:28 & 2279 & GCS 12 & 11.6 & 266.33030 & -28.67280 & 283.8 & \nodata & 9 & 0.2 \\ 
2007 Feb 24 04:56:00 & 7047 & Deep GCS 26 & 36.7 & 266.31970 & -28.66905 & 90.8 & 9 & \nodata & 0.17 \\ [2pt]
2007 Jul 19 10:04:05 & 7046 & Deep GCS 25 & 36.5 & 266.82943 & -28.72059 & 278.4 & 18 & \nodata & 0.10 \\ 
2001 Jul 17 14:11:51 & 2288 & GCS 7 & 11.1 & 266.82550 & -28.70883 & 283.8 & \nodata & 10 & 0.2 \\
2006 Nov 01 14:11:44 & 7044 & Deep GCS 23 & 37.9 & 266.63030 & -28.61480 & 268.2 & 12 & \nodata & 0.11 \\
2001 Jul 17 17:51:28 & 2292 & GCS 8 & 11.6 & 266.63554 & -28.60780 & 283.8 & \nodata & 12 & 0.2 \\
2001 Jul 17 21:19:58 & 2294 & GCS 9 & 11.6 & 266.44598 & -28.50635 & 283.8 & \nodata & 8 & 0.2 \\  [2pt]
2007 Feb 23 07:15:57 & 7045 & Deep GCS 24 & 37.0 & 266.43720 & -28.50006 & 90.9 & 9 & \nodata & 0.13 \\
2001 Jul 16 11:52:55 & 2277 & GCS 4 & 10.4 & 266.94060 & -28.54206 & 283.8 & \nodata & \nodata & 0.5 \\
2000 Mar 29 09:44:36 & 944  & Sgr B2 & 97.5 & 266.78070 & -28.44160 & 87.8 & 23 & \nodata & 0.10 \\ 
2001 Jul 16 15:01:25 & 2280 & GCS 5 & 10.4 & 266.75080 & -28.44106 & 283.8 & \nodata & 9 & 0.2 \\
2001 Jul 16 18:09:55 & 2285 & GCS 6 & 10.4 & 266.56137 & -28.33985 & 283.4 & \nodata & \nodata & 0.5 \\ [2pt]
2001 Jul 16 02:15:50 & 2269 & GCS 1 & 10.5 & 267.05519 & -28.37520 & 283.8 & \nodata & \nodata &  0.5 \\ 
2001 Jul 16 05:35:55 & 2271 & GCS 2 & 10.4 & 266.86561 & -28.27427 & 283.8 & \nodata & \nodata & 0.5 \\
2001 Jul 16 08:44:25 & 2274 & GCS 3 & 10.4 & 266.67640 & -28.17316 & 283.8 & \nodata & \nodata & 0.5 
\enddata
\tablecomments{The columns are: the date and time of the observation (UT);
the observation identifier; the exposure time (ks); the right ascension and 
declination (J2000), the roll angle of the satellite; the number of 2MASS 
sources that we used to align the astrometry for the deppest observations 
at any point; the number of X-ray matches that we used to align the 
astrometry of the shallower observations to the deepest ones; and the 
1$\sigma$ uncertainties on the astrometry as determined from the standard 
deviations in the means of the offsets between the input and trial catalogs. 
The observations 
have been sorted first by declination, then by right ascension, so that 
observations of the same region are grouped in the table.}
\end{deluxetable*}
%% sources within 5\arcmin\ detected at $>$3$\sigma$. Additionally, 
%% for IR matching, required the sources be detected in the 0.5-2.0 keV 
%% band. 

\clearpage

\begin{landscape}
  
\begin{deluxetable}{lccccccccccccccc}
%\tabletypesize{\scriptsize}
%\rotate
\tablecolumns{16}
\tablewidth{0pc}
\tablecaption{Galactic Center X-ray Source Locations and Extraction Information\label{tab:positions}}
\tablehead{
\colhead{Number} & \colhead{Name} & 
\colhead{RA} & \colhead{Dec} & \colhead{$\sigma_{\rm X}$} &
\colhead{Pos.} & \colhead{Field} & \colhead{Band} & \colhead{Offset} & 
\colhead{$N_{\rm Obs}$} & \colhead{Exposure} & \colhead{$f_{\rm PSF}$} &
\colhead{$E_{\rm PSF}$} & \colhead{$R_{\rm src}$} & \colhead{$F_{\rm sens}$} &
\colhead{Flags} \\
%% line 2
\colhead{} & \colhead{(CXOUGC J)} & 
\multicolumn{2}{c}{(degrees, J2000)} & \colhead{(arcsec)} &
\colhead{} & \colhead{} & \colhead{} & \colhead{(arcmin)} & 
\colhead{} & \colhead{(s)} & \colhead{} &
\colhead{(keV)} & \colhead{(arcsec)} & \colhead{($10^{-7}$ cm$^{-2}$ s$^{-1}$)} &
\colhead{} \\
%% line 3
\colhead{1} & \colhead{2} & 
\colhead{3} & \colhead{4} & \colhead{5} &
\colhead{6} & \colhead{7} & \colhead{8} & \colhead{9} & 
\colhead{10} & \colhead{11} & \colhead{12} &
\colhead{13} & \colhead{14} & 
\colhead{15} & \colhead{16}
}
\startdata
    1 & 174457.1--285740 & 266.23813 & $--28.96121$ &  1.8 & c & full  & soft &  9.8 & 14 &  372669 & 0.90 & 1.50 &  9.8 &  32.2 &    s    \\
    2 & 174457.4--285622 & 266.23941 & $--28.93967$ &  2.0 & c & 2004  & full & 10.2 & 15 &  377775 & 0.90 & 4.51 & 11.2 &  26.8 &    lb   \\
    3 & 174459.1--290604 & 266.24620 & $--29.10122$ &  1.3 & c & full  & full & 10.8 & 12 &  571227 & 0.87 & 4.51 & 11.8 &   9.0 &    sc   \\
    4 & 174459.9--290324 & 266.24982 & $--29.05683$ &  2.0 & c & 2002  & full &  9.3 & 23 &  923358 & 0.90 & 4.51 &  9.6 &  10.0 &    g    \\
    5 & 174459.9--290538 & 266.24994 & $--29.09415$ &  1.5 & c & full  & hard & 10.3 & 18 &  732385 & 0.90 & 4.51 & 11.5 &   9.6 &    b    \\ [2pt]
    6 & 174500.2--290057 & 266.25113 & $--29.01598$ &  2.6 & w & 2005  & soft &  8.6 & 18 &  423683 & 0.76 & 1.50 &  6.3 &  22.7 &    c    \\
    7 & 174500.6--290443 & 266.25244 & $--29.07895$ &  1.3 & c & full  & full &  9.7 & 20 &  810677 & 0.69 & 4.51 &  6.8 &  10.2 &    lc   \\
    8 & 174501.3--285501 & 266.25580 & $--28.91719$ &  0.5 & w & 242   & full & 10.1 & 18 &  423686 & 0.77 & 4.51 &  7.9 &  23.3 &   lcp   \\
    9 & 174501.4--290408 & 266.25598 & $--29.06908$ &  1.1 & c & full  & hard &  9.3 & 26 &  945772 & 0.80 & 4.51 &  7.6 &  10.2 &    c    \\
   10 & 174501.7--290313 & 266.25732 & $--29.05367$ &  1.1 & c & full  & hard &  8.9 & 27 &  956403 & 0.90 & 4.51 &  9.3 &  11.5 &    gb   \\ [2pt]
   11 & 174501.8--290206 & 266.25760 & $--29.03520$ &  1.1 & c & full  & full &  8.6 & 26 &  943948 & 0.69 & 4.51 &  5.3 &  10.8 &    c    \\
   12 & 174501.9--285719 & 266.25827 & $--28.95553$ &  1.8 & c & full  & soft &  8.9 & 18 &  423686 & 0.69 & 1.50 &  5.1 &  24.7 &    lbc  \\
   13 & 174502.2--285749 & 266.25946 & $--28.96381$ &  1.0 & c & 5951  & full &  8.7 & 18 &  423686 & 0.87 & 4.51 &  7.5 &  23.5 &    c    \\
   14 & 174502.4--290205 & 266.26007 & $--29.03492$ &  1.1 & c & 2002  & soft &  8.5 & 27 &  956403 & 0.70 & 1.50 &  4.9 &   9.7 &    c    \\
   15 & 174502.4--290453 & 266.26039 & $--29.08160$ &  1.4 & c & full  & soft &  9.4 & 29 &  967611 & 0.88 & 1.50 &  8.4 &  11.4 &   -     \\ [2pt]
   16 & 174502.5--290415 & 266.26077 & $--29.07086$ &  2.0 & w & full  & tile &  9.1 & 29 &  967611 & 0.75 & 4.51 &  6.5 &  11.2 &    lbc  \\
   17 & 174502.7--290127 & 266.26101 & $--29.02376$ &  1.6 & c & 3392  & full &  8.3 & 27 &  956431 & 0.89 & 4.51 &  7.7 &   9.2 &   -     \\
   18 & 174502.8--290429 & 266.26198 & $--29.07480$ &  1.3 & c & full  & soft &  9.1 & 30 &  972102 & 0.77 & 1.50 &  6.1 &  11.1 &    bc   \\
   19 & 174502.9--285920 & 266.26222 & $--28.98968$ &  1.1 & c & full  & full &  8.2 & 18 &  423686 & 0.90 & 4.51 &  7.3 &  19.4 &   -     \\
   20 & 174503.7--285805 & 266.26531 & $--28.96845$ &  1.2 & c & 2005  & full &  8.3 & 20 &  440908 & 0.70 & 4.51 &  4.8 &  21.7 &    bc   \\ [2pt]
   21 & 174503.8--290004 & 266.26584 & $--29.00112$ &  0.9 & c & full  & full &  8.0 & 30 &  985035 & 0.70 & 4.51 &  4.5 &   9.5 &    gc   \\
   22 & 174504.1--285902 & 266.26740 & $--28.98400$ &  4.7 & w & 6646  & soft &  8.0 & 30 &  955147 & 0.74 & 1.50 &  4.7 &  13.8 &    bc   \\
   23 & 174504.2--285653 & 266.26758 & $--28.94817$ &  1.4 & c & 242   & full &  8.6 & 22 &  457849 & 0.90 & 4.51 &  7.8 &  22.7 &    l    \\
   24 & 174504.2--290410 & 266.26764 & $--29.06977$ &  1.5 & c & 2002  & soft &  8.7 & 34 & 1011810 & 0.70 & 1.50 &  4.9 &   8.7 &    c    \\
   25 & 174504.2--290610 & 266.26841 & $--29.10258$ &  1.9 & c & 2005  & hard &  9.8 & 32 &  988816 & 0.83 & 4.51 &  8.5 &  10.3 &    c    \\ 
\enddata
\tablecomments{A portion of the full table is shown here, for guidance as 
to its form and content. The columns are described in the text. 
}
\end{deluxetable}

\clearpage

\begin{deluxetable}{lcccccccccccccc}
%\tabletypesize{\scriptsize}
%\rotate
\tablecolumns{15}
\tablewidth{0pc}
\tablecaption{Galactic Center X-ray Source Photometry\label{tab:phot}}
\tablehead{
\colhead{No.} & \colhead{Name} & 
\colhead{$\log(P_{\rm KS}$)} &
\colhead{$C_{\rm t,0.5-2}$} & \colhead{$C_{\rm b,0.5-2}$} & 
\colhead{$C_{\rm net,0.5-2}$} &
\colhead{$C_{\rm t,2-8}$} & \colhead{$C_{\rm b,2-8}$} & 
\colhead{$C_{\rm net,2-8}$} &
\colhead{$F_{0.5-2}$} & \colhead{$F_{2-8}$} & \colhead{$\bar{E}$} & 
\colhead{$HR_0$} & \colhead{$HR_1$} & \colhead{$HR_2$} \\
% line 2
\colhead{} & \colhead{CXOUGC J} & \colhead{} &
\colhead{} & \colhead{} & 
\colhead{} &
\colhead{} & \colhead{} & 
\colhead{} &
\multicolumn{2}{c}{($10^{-7}$ cm$^{-2}$ s$^{-1}$)} & \colhead{} & 
\colhead{} & \colhead{} & \colhead{} \\
%line 3
\colhead{1} & \colhead{2} & 
\colhead{3} &
\colhead{4} & \colhead{5} & 
\colhead{6} &
\colhead{7} & \colhead{8} & 
\colhead{9} &
\colhead{10} & \colhead{11} & 
\colhead{12} & \colhead{13} & \colhead{14} & \colhead{15}
}
\startdata
    1 & 174457.1--285740 &  --4.50 &      86 &   26.3 & $    59.7_{-  14.0}^{+  16.4}$ &    185 &  186.4 & $<  23.7$ &     6.8 & \nodata & 1.0 & $<-0.555$ & \nodata & \nodata \\
    2 & 174457.4--285622 &  --0.09 &      52 &   53.0 & $<  14.2$ &    357 &  298.8 & $    58.2_{-  28.3}^{+  33.6}$ & \nodata &     8.2 & 4.6 & \nodata & $> 0.031$ & $-0.169_{- 0.615}^{+ 0.612}$ \\
    3 & 174459.1--290604 &  --2.57 &     118 &   88.1 & $    29.9_{-  16.7}^{+  18.8}$ &    498 &  349.1 & $   148.9_{-  40.4}^{+  32.3}$ &     1.9 &    13.3 & 3.5 & $ 0.116_{- 0.411}^{+ 0.415}$ & $ 0.299_{- 0.263}^{+ 0.294}$ & $-0.254_{- 0.396}^{+ 0.254}$ \\
    4 & 174459.9--290324 &  --1.09 &      96 &   91.1 & $<  15.0$ &    369 &  334.6 & $    34.4_{-  28.9}^{+  29.6}$ & \nodata &     2.4 & 3.1 & $>-0.066$ & $< 0.194$ & \nodata \\
    5 & 174459.9--290538 &  --0.97 &     107 &  115.0 & $<  22.7$ &    550 &  434.9 & $   115.1_{-  40.2}^{+  40.2}$ & \nodata &    10.3 & 4.8 & \nodata & $> 0.243$ & $-0.012_{- 0.374}^{+ 0.357}$ \\ [2pt]
    6 & 174500.2--290057 &  --1.99 &      41 &   25.1 & $    15.9_{-   9.5}^{+  11.4}$ &    138 &  133.5 & $<  19.2$ &     1.5 & \nodata & 0.5 & $-0.065_{- 0.779}^{+ 0.506}$ & $< 0.006$ & \nodata \\
    7 & 174500.6--290443 &  --1.90 &      50 &   44.5 & $<  10.8$ &    316 &  200.3 & $   115.7_{-  27.0}^{+  31.5}$ & \nodata &    11.1 & 4.6 & \nodata & $> 0.400$ & $ 0.053_{- 0.246}^{+ 0.247}$ \\
    8 & 174501.3--285501 & --40.35 &     175 &   69.8 & $   105.2_{-  21.0}^{+  22.4}$ &  13585 & 3163.7 & $ 10421.3_{- 165.9}^{+ 165.9}$ &     9.7 &  1625.1 & 4.7 & $ 0.861_{- 0.028}^{+ 0.027}$ & $ 0.462_{- 0.021}^{+ 0.021}$ & $ 0.146_{- 0.017}^{+ 0.017}$ \\
    9 & 174501.4--290408 &  --4.12 &      92 &  100.1 & $<  21.7$ &    523 &  371.2 & $   151.8_{-  38.3}^{+  38.3}$ & \nodata &    11.9 & 5.4 & \nodata & $> 0.029$ & $ 0.578_{- 0.230}^{+ 0.242}$ \\
   10 & 174501.7--290313 &  --0.76 &      89 &  104.1 & $<  25.7$ &    561 &  433.0 & $   128.0_{-  40.4}^{+  40.4}$ & \nodata &     8.7 & 5.1 & $> 0.390$ & $ 0.163_{- 0.547}^{+ 0.596}$ & $ 0.351_{- 0.335}^{+ 0.362}$ \\ [2pt]
   11 & 174501.8--290206 &  --2.37 &      39 &   34.2 & $<  10.1$ &    248 &  144.9 & $   103.1_{-  24.1}^{+  27.7}$ & \nodata &     8.0 & 4.6 & \nodata & $> 0.440$ & $-0.058_{- 0.253}^{+ 0.244}$ \\
   12 & 174501.9--285719 &  --0.88 &      41 &   19.9 & $    21.1_{-   9.7}^{+  11.2}$ &    163 &  143.6 & $    19.4_{-  18.1}^{+  18.6}$ &     2.1 &     3.2 & 3.2 & $<-0.296$ & $>-0.014$ & $< 0.288$ \\
   13 & 174502.2--285749 &  --5.62 &      60 &   38.7 & $    21.3_{-  12.0}^{+  13.3}$ &    378 &  281.3 & $    96.7_{-  27.7}^{+  36.2}$ &     1.6 &     6.6 & 2.6 & $ 0.517_{- 0.238}^{+ 0.257}$ & $-0.221_{- 0.259}^{+ 0.259}$ & $<-0.564$ \\
   14 & 174502.4--290205 &  --2.76 &      70 &   31.1 & $    38.9_{-  14.5}^{+  12.5}$ &    159 &  109.7 & $    49.3_{-  21.9}^{+  19.1}$ &     1.8 &     3.2 & 2.8 & $-0.570_{- 0.417}^{+ 0.288}$ & $ 0.417_{- 0.391}^{+ 0.568}$ & $-0.339_{- 0.640}^{+ 0.409}$ \\
   15 & 174502.4--290453 &  --7.28 &     172 &   91.6 & $    80.4_{-  19.0}^{+  24.0}$ &    349 &  340.4 & $<  27.8$ &     3.3 & \nodata & 0.9 & $-0.577_{- 0.333}^{+ 0.233}$ & $<-0.064$ & \nodata \\ [2pt]
   16 & 174502.5--290415 &  --0.04 &      69 &   56.4 & $    12.6_{-  12.5}^{+  10.5}$ &    218 &  194.8 & $<  15.4$ &     0.6 & \nodata & 3.1 & $ 0.071_{- 0.918}^{+ 0.948}$ & $<-0.149$ & \nodata \\
   17 & 174502.7--290127 &  --1.10 &      80 &  101.9 & $<  30.8$ &    372 &  312.8 & $    59.2_{-  28.8}^{+  34.1}$ & \nodata &     3.4 & 6.3 & \nodata & $>-0.097$ & $ 0.161_{- 0.683}^{+ 0.550}$ \\
   18 & 174502.8--290429 &  --0.49 &     139 &   86.3 & $    52.7_{-  20.0}^{+  18.3}$ &    308 &  279.6 & $<  18.9$ &     2.2 & \nodata & 2.9 & $<-0.609$ & \nodata & $> 0.065$ \\
   19 & 174502.9--285920 &  --1.28 &      43 &   33.7 & $<   7.5$ &    234 &  164.0 & $    70.0_{-  23.2}^{+  27.0}$ & \nodata &     9.0 & 4.9 & \nodata & $> 0.581$ & $ 0.084_{- 0.294}^{+ 0.323}$ \\
   20 & 174503.7--285805 &  --0.63 &      17 &   15.3 & $<   6.9$ &    143 &   75.0 & $    68.0_{-  19.6}^{+  19.4}$ & \nodata &    11.2 & 4.7 & $> 0.181$ & $ 0.210_{- 0.445}^{+ 0.596}$ & $ 0.320_{- 0.294}^{+ 0.312}$ \\ [2pt]
   21 & 174503.8--290004 &  --1.94 &      56 &   51.9 & $<  12.8$ &    341 &  232.2 & $   108.8_{-  30.2}^{+  30.3}$ & \nodata &     6.8 & 4.3 & $> 0.108$ & $ 0.418_{- 0.293}^{+ 0.436}$ & $-0.165_{- 0.322}^{+ 0.292}$ \\
   22 & 174504.1--285902 &  --0.22 &      27 &   19.9 & $<   6.9$ &     90 &   71.8 & $    18.2_{-  14.1}^{+  15.4}$ & \nodata &     1.2 & 2.7 & $>-0.282$ & $ 0.228_{- 0.716}^{+ 0.767}$ & $<-0.148$ \\
   23 & 174504.2--285653 &  --2.59 &      46 &   56.2 & $<  18.2$ &    419 &  353.4 & $    65.6_{-  32.1}^{+  34.7}$ & \nodata &    10.3 & 6.5 & \nodata & \nodata & $> 0.435$ \\
   24 & 174504.2--290410 &  --2.66 &     121 &   66.9 & $    54.1_{-  19.1}^{+  16.6}$ &    250 &  249.8 & $<  27.0$ &     2.1 & \nodata & 0.9 & $<-0.518$ & \nodata & \nodata \\
   25 & 174504.2--290610 &  --1.31 &     129 &  124.1 & $<  19.0$ &    543 &  496.0 & $    47.0_{-  41.3}^{+  41.3}$ & \nodata &     3.1 & 4.6 & \nodata & $> 0.406$ & $-0.140_{- 0.851}^{+ 0.485}$ \\ 
\enddata
\tablecomments{A portion of the full table is shown here, for guidance as 
to its form and content. The columns are described in the text}
\end{deluxetable}

\clearpage
\end{landscape}

\begin{deluxetable*}{llcccccc}[htp]
\tabletypesize{\scriptsize}
\tablecolumns{8}
\tablewidth{0pc}
\tablecaption{Sources with Long-term Variability\label{tab:longterm}}
\tablehead{
\colhead{Number} & \colhead{Name} & \colhead{Location} & 
\colhead{Min. ObsID} & \colhead{$F_{\rm min}$} & 
\colhead{Max. Obsid} & \colhead{$F_{\rm max}$} & 
\colhead{$F{\rm max}/F_{\rm min}$}\\
%% line 2
\colhead{} & \colhead{(CXOUGC J)} & \colhead{} &
\colhead{} & \colhead{($10^{-7}$ cm$^{-2}$ s$^{-1}$)} &
\colhead{} & \colhead{($10^{-7}$ cm$^{-2}$ s$^{-1}$)} & 
\colhead{} 
}
\startdata
   2 & 174457.4-285622 & gc & 5953 & $< 10$ & 4684 & $    31$$\pm$$  8$ & $>    2.2$ \\
   7 & 174500.6-290443 & gc & 7556 & $< 27$ & 3392 & $    14$$\pm$$  4$ & $>    0.4$ \\
   8 & 174501.3-285501 & gc & 4684 & $    8$$\pm$$  6$ & 1561a & $12437$$\pm$$166$ & $1530_{- 676}^{+5824}$ \\
  12 & 174501.9-285719 & f & 2284 & $< 10$ & 5953 & $    16$$\pm$$  8$ & $>    0.8$ \\
  16 & 174502.5-290415 & gc & 2943 & $<  5$ & 7556 & $    64$$\pm$$ 41$ & $>    4.2$ \\ [2pt]
  23 & 174504.2-285653 & gc & 4683 & $<  6$ & 5953 & $    16$$\pm$$  9$ & $>    1.1$ \\
  31 & 174504.8-285410 & f & 5951 & $< 16$ & 6363 & $    38$$\pm$$ 14$ & $>    1.5$ \\
  35 & 174505.2-285713 & f & 2953 & $< 25$ & 4683 & $    16$$\pm$$  7$ & $>    0.4$ \\
  45 & 174506.1-285710 & gc & 5954 & $<  6$ & 2953 & $    26$$\pm$$ 21$ & $>    0.9$ \\
  53 & 174507.0-290452 & gc & 6113 & $< 31$ & 2953 & $    29$$\pm$$ 20$ & $>    0.3$ \\ [2pt]
  54 & 174507.1-285720 & gc & 5950 & $<  6$ & 2951 & $    21$$\pm$$ 15$ & $>    1.0$ \\
  58 & 174507.5-285614 & f & 3665 & $< 12$ & 6640 & $    47$$\pm$$ 44$ & $>    0.3$ \\
  71 & 174508.4-290033 & f & 3663 & $<  5$ & 5360 & $   375$$\pm$$ 67$ & $>   61.1$ \\
  80 & 174509.2-285457 & gc & 2951 & $<  9$ & 5954 & $    20$$\pm$$ 13$ & $>    0.8$ \\
  89 & 174509.5-285502 & gc & 3392 & $<  3$ & 6363 & $    19$$\pm$$ 10$ & $>    3.1$ \\ [2pt]
  96 & 174510.1-285624 & f & 3549 & $<  6$ & 1561a & $    16$$\pm$$ 12$ & $>    0.7$ \\
  98 & 174510.2-285505 & gc & 3663 & $<  5$ & 6642 & $    51$$\pm$$ 40$ & $>    2.3$ \\
 100 & 174510.3-290642 & gc & 7557 & $< 24$ & 2954 & $    55$$\pm$$ 23$ & $>    1.3$ \\
 101 & 174510.4-285433 & gc & 4683 & $<  5$ & 5953 & $    19$$\pm$$  8$ & $>    2.5$ \\
 102 & 174510.4-285544 & gc & 5950 & $<  5$ & 3665 & $    11$$\pm$$  4$ & $>    1.6$ \\ [2pt]
 103 & 174510.4-285545 & gc & 5950 & $<  5$ & 3665 & $    12$$\pm$$  4$ & $>    1.6$ \\
 106 & 174510.6-285437 & gc & 3549 & $<  7$ & 3663 & $    14$$\pm$$  8$ & $>    0.9$ \\
 109 & 174510.8-285606 & gc & 6641 & $< 22$ & 3665 & $    12$$\pm$$  4$ & $>    0.4$ \\
 112 & 174510.9-285508 & gc & 5954 & $<  9$ & 4683 & $    18$$\pm$$  7$ & $>    1.3$ \\
 129 & 174511.6-285915 & gc & 5953 & $<  3$ & 3392 & $     8$$\pm$$  2$ & $>    1.6$ \\ [2pt]
 143 & 174512.1-290005 & gc & 2952 & $<  7$ & 3393 & $    15$$\pm$$  3$ & $>    1.6$ \\
 155 & 174512.4-285318 & gc & 7555 & $< 27$ & 3393 & $    11$$\pm$$  4$ & $>    0.3$ \\
 164 & 174512.8-285441 & gc & 6644 & $< 22$ & 2953 & $    33$$\pm$$ 20$ & $>    0.6$ \\
 191 & 174514.1-285426 & gc & 5953 & $<  7$ & 5954 & $    55$$\pm$$ 20$ & $>    4.7$ \\
 226 & 174515.1-290006 & gc & 4684 & $<  3$ & 3392 & $    10$$\pm$$  2$ & $>    2.8$
\enddata
\tablecomments{The columns of the table are: the record number from
Table~2, the source name, a flag stating whether a source is in the 
foreground or near the Galactic center, the ObsID in which the 
minimum flux was observed, the minimum flux, the ObsID in which 
the maximum flux was observed, the maximum flux, and the ratio
of the maximum to minimum fluxes. 
The first 30 lines are shown as an example; the full table is accessible 
on-line.}
\end{deluxetable*}

\clearpage

\begin{deluxetable*}{llcccccccc}[htp]
\tabletypesize{\scriptsize}
\tablecolumns{10}
\tablewidth{0pc}
\tablecaption{Sources with Short-term Variability\label{tab:shortterm}}
\tablehead{
\colhead{Number} & \colhead{Name} & \colhead{Location} & 
\colhead{ObsID} & \colhead{Prior Odds} & \colhead{$N_{\rm blocks}$} & 
\colhead{$T_{\rm peak}$} & \colhead{$F_{\rm min}$} & \colhead{$F_{\rm max}$} & 
\colhead{$F{\rm max}/F_{\rm min}$}\\
%% line 2
\colhead{} & \colhead{(CXOUGC J)} & \colhead{} &
\colhead{} & \colhead{} & \colhead{} &
\colhead{(s)} & \multicolumn{2}{c}{($10^{-7}$ \phcms)} & 
\colhead{} 
}
\startdata
   1 & 174457.1-285740 & f & 1561b &  100 &  2 &   3152 & $< 17$ & $  193^{+   72}_{-   55}$ & $>   3.3$ \\
   3 & 174459.1-290604 & gc & 3663 &   10 &  2 &  19600 & $< 11$ & $   48^{+   17}_{-   15}$ & $>   1.4$ \\
  71 & 174508.4-290033 & f & 5360 & 1000 &  2 &   2736 & $< 63$ & $  673^{+  119}_{-  103}$ & $>   1.6$ \\
 186 & 174513.7-285638 & gc & 4684 &   10 &  2 &  21312 & $<  7$ & $   36^{+   12}_{-    9}$ & $>   1.4$ \\
     257 & 174516.6-285412 & f & 4684 & 1000 &  2 &  16512 & $    5^{+  7}_{-  4}$ & $   90^{+  23}_{-  19}$ & $  20.0\pm   24.4$ \\ [2pt]
     304 & 174517.8-290653 & gc & 3392 & 1000 &  2 & 110160 & $    4^{+  5}_{-  4}$ & $   40^{+   6}_{-   5}$ & $   9.6\pm   10.5$ \\
     364 & 174519.5-285955 & gc & 4683 &   10 &  2 &   6896 & $    3^{+  3}_{-  2}$ & $   57^{+  26}_{-  19}$ & $  18.2\pm   18.0$ \\
     398 & 174520.3-290143 & gc & 2943 & 1000 &  2 &   3472 & $    5^{+  5}_{-  3}$ & $  134^{+  58}_{-  42}$ & $  29.6\pm   29.0$ \\
     424 & 174520.6-290152 & f & 3392 & 1000 &  5 &  33360 & $   34^{+  6}_{-  5}$ & $  526^{+  35}_{-  33}$ & $  15.3\pm    2.7$ \\
     424 &   &   & 3393 & 1000 &  4 &  22144 & $   15^{+  3}_{-  3}$ & $  858^{+  55}_{-  52}$ & $  56.1\pm   11.0$ \\ [2pt]
     424 &   &   & 5951 & 1000 &  2 &  23760 & $    9^{+  8}_{-  5}$ & $  104^{+  17}_{-  15}$ & $  11.3\pm    8.2$ \\
     470 & 174521.7-290151 & gc & 5950 &  100 &  2 &  10400 & $    2^{+  4}_{-  2}$ & $   52^{+  19}_{-  15}$ & $  22.9\pm   30.4$ \\
     472 & 174521.8-285912 & f & 3392 & 1000 &  3 &  18560 & $    3^{+  3}_{-  2}$ & $  103^{+  17}_{-  15}$ & $  30.0\pm   21.6$ \\
     663 & 174525.1-285703 & f & 3665 & 1000 &  2 &   1664 & $   12^{+  4}_{-  3}$ & $  566^{+ 176}_{- 137}$ & $  49.0\pm   20.6$ \\
 811 & 174527.1-290730 & gc & 3549 &   10 &  2 &   3296 & $<  6$ & $   87^{+   57}_{-   38}$ & $>   5.9$ \\ [2pt]
    1100 & 174530.3-290341 & gc & 3392 & 1000 &  3 &  21440 & $    5^{+  8}_{-  5}$ & $   91^{+  15}_{-  13}$ & $  17.3\pm   20.5$ \\
    1100 &   &   & 5950 & 1000 &  2 &  17120 & $   55^{+  9}_{-  8}$ & $  136^{+  20}_{-  17}$ & $   2.5\pm    0.5$ \\
    1183 & 174531.1-290219 & gc & 3393 &   10 &  2 &  70704 & $    8^{+  3}_{-  2}$ & $   25^{+   4}_{-   4}$ & $   2.9\pm    1.0$ \\
1525 & 174534.2-290011 & f & 3392 &  100 &  2 &  49392 & $<  1$ & $   14^{+    4}_{-    3}$ & $>   2.7$ \\
1569 & 174534.5-290236 & gc & 3392 & 1000 &  2 &  73504 & $<  1$ & $   10^{+    3}_{-    2}$ & $>   2.4$ \\ [2pt]
1608 & 174534.8-290851 & f & 1561a &  100 &  2 &  21168 & $< 11$ & $   78^{+   17}_{-   14}$ & $>   1.3$ \\
    1676 & 174535.5-290124 & gc & 3549 & 1000 &  2 &  15920 & $ 1234^{+ 77}_{- 73}$ & $ 1877^{+  71}_{-  68}$ & $   1.5\pm    0.1$ \\
    1676 &   &   & 5950 &  100 &  2 &  16544 & $  156^{+ 14}_{- 13}$ & $  267^{+  27}_{-  24}$ & $   1.7\pm    0.2$ \\
    1676 &   &   & 6644 & 1000 &  2 &   2304 & $  179^{+ 78}_{- 57}$ & $  809^{+ 155}_{- 131}$ & $   4.5\pm    1.9$ \\
    1686 & 174535.6-290133 & gc & 3665 & 1000 &  3 &  61632 & $   14^{+  6}_{-  4}$ & $  178^{+  15}_{-  14}$ & $  12.9\pm    5.0$ \\ [2pt]
    1686 &   &   & 6641 & 1000 &  2 &   4336 & $  650^{+238}_{-183}$ & $ 5389^{+ 222}_{- 213}$ & $   8.3\pm    2.7$ \\
    1691 & 174535.7-285357 & f & 5951 &  100 &  2 &  29264 & $   10^{+ 11}_{-  8}$ & $   79^{+  13}_{-  12}$ & $   7.9\pm    7.5$ \\
1706 & 174535.8-290159 & gc & 3393 & 1000 &  2 &  90464 & $<  3$ & $   14^{+    3}_{-    3}$ & $>   0.8$ \\
    1748 & 174536.1-290806 & gc & 3393 & 1000 &  2 &  75104 & $    3^{+  4}_{-  3}$ & $   31^{+   6}_{-   5}$ & $   8.9\pm    9.3$ \\
    1765 & 174536.3-285545 & f & 3392 &  100 &  4 &   7568 & $    3^{+  2}_{-  1}$ & $  252^{+  81}_{-  63}$ & $  94.5\pm   60.5$
\enddata
\tablecomments{The columns of the table are: the record number from
Table~2, the source name, a flag stating whether a source is in the 
foreground or near the Galactic center, the ObsID in which the 
variability was identified, the odds ratio used as a prior in the
Bayesian Blocks routine when characterizing the variability, the 
number of blocks used to describe the data, the duration of the block
with the maximum flux, the minimum flux, the maximum flux, and the ratio
of the maximum to minimum fluxes. The table only includes variable sources 
that were successfully characterized by the Bayesian Blocks routine. 
The first 30 lines are shown as an example; the full table is accessible 
on-line.}
\end{deluxetable*}

\end{document}